\documentclass[aip,jcp,reprint,groupedaddress,showpacs]{revtex4-1}
\usepackage[T1]{fontenc}
\usepackage[utf8]{inputenc}
\usepackage{amsmath}
\usepackage{graphicx}
\usepackage{amssymb}
\usepackage{amsmath}
\usepackage{pstricks}
\usepackage{textcomp}
\usepackage{lineno}

\usepackage[normalem]{ulem}

\begin{document}
\title{Shortcomings of the Bond Orientational Order Parameters for the Analysis of Disordered Particulate Matter\vspace*{0.2cm}}

\author{Walter~Mickel$^{1,2,3}$}
\email{Walter.Mickel@kit.edu}
\author{Sebastian C.~Kapfer$^{1}$}
\email{Sebastian.Kapfer@physik.fau.de}
\author{Gerd~E.~Schr\"oder-Turk$^1$}
\email{Gerd.Schroeder-Turk@physik.fau.de}
\author{Klaus~Mecke$^1$}
\email{Klaus.Mecke@physik.fau.de}
\affiliation{\mbox{\vspace*{0.2cm}$^1$ Theoretische Physik,
Friedrich-Alexander-Universit\"at Erlangen, Staudtstr.~7, D-91058
Erlangen, Germany}}
\affiliation{\mbox{$^2$ Universit\'{e} de Lyon, F-69000, Lyon, France and CNRS,
UMR5586, Laboratoire PMCN, Lyon, France}}
\affiliation{\mbox{$^3$ Institute for
Stochastics, Karlsruhe Institute of Technology, D-76128 Karlsruhe, Germany}}
\date{\today}

\begin{abstract}
Local structure characterization with the bond-orientational order
parameters $q_4$, $q_6$,$\ldots$ introduced by Steinhardt {\it et al.}~has
become a standard tool in condensed matter physics,
with applications including glass, jamming, melting or crystallization
transitions and cluster formation.
Here we discuss two fundamental flaws in the definition of these parameters that
significantly affect their interpretation for studies of disordered systems, and
offer a remedy.
First, the definition of the bond-orientational order parameters considers
the geometrical arrangement of a set of neighboring spheres $\mathrm{NN}(p)$
around a given central particle $p$;
we show that procedure to select the spheres constituting the neighborhood
$\mathrm{NN}(p)$ can have greater influence on both the numerical values and
qualitative trend of $q_l$ than a change of the physical parameters, such as
packing fraction.
Second, the discrete nature of neighborhood implies that
$\mathrm{NN}(p)$ is not a continuous function of the particle coordinates; this
discontinuity, inherited by $q_l$, leads to a lack of robustness of the $q_l$ as
structure metrics.
Both issues can be avoided by a morphometric approach leading to the robust
\emph{Minkowski structure metrics} $q_l'$.  These $q_l'$ are of a similar mathematical form as the conventional bond-orientational order
parameters and are mathematically
equivalent to the recently introduced Minkowski tensors
[Europhys.~Lett.~{\bf 90}, 34001 (2010); Phys.~Rev.~E.~{\bf 85}, 030301 (2012)].\\

\end{abstract}

\maketitle

In 1983 Steinhardt \emph{et al.}~\cite{PRB83_Steinhardt} proposed
the family of local $q_l$ and global $Q_l$ bond-orientational order (BOO)
parameters as a three-dimensional generalization of the $\psi_6$ hexatic order parameter in
two dimensions \cite{PRB79_Nelson}.
Bond orientation analysis has become the most commonly 
used tool for the identification of  different crystalline phases and clusters,
notably fcc, hcp and bcc \cite{PRL95_tenWolde,JChemPhys11_Ni, EPJE10_Xu,
JChemPhys08_Lechner, JChemPhys09_Valdes, JPCM10_Kawasaki, PRE07_Wang} or
icosahedral nuclei \cite{JChemPhys05_Wang, PRL07_Keys, PRE07_Iacovella}. They
are also used to study melting transitions \cite{JChemPhys05_Wang,
JChemPhys07_Chakravarty, JChemPhys09_Calvo, JChemPhys07_Chakravarty} 
and interfaces in colloidal fluids and crystals \cite{PNAS09-Hernandez-Guzman}.
For the study of glasses and super-cooled fluids $q_6$ and $Q_6$
have become the most prominent order parameter when searching
for glass transitions \cite{KobBinder,PRL11_Ikeda, JChemPhys09_Mokshin, NatMat10_Tanaka}
and crystalline clusters \cite{EPJB06_Lochmann, JChemPhys11_Ni, PRL10_Schilling,
PRL07_Keys, JChemPhys91_vanDuijneveldt, JPCM10_Kawasaki}. 
While $q_l$ is defined as a local parameter for each particle, 
other studies have used global averages of bond angles
($Q_l$) to detect single-crystalline order across the entire sample
\cite{JChemPhys06_Wouterse,PRE07_Duff,PRE08_Abraham}. 

The BOO parameters $q_l$ and $Q_l$ are defined as structure metrics for ensembles
of $N$ spherical particles.
For a given sphere $a$ one assigns a set of nearest neighbors (NN) spheres
$\mathrm{NN}(a)$. The number of NN assigned to $a$ is $n(a)=|\mathrm{NN}(a)|$.
Any two spheres $a$ and $b$ are said to be connected by a {\em bond} if they are
neighbors, i.e.~if $a\in \mathrm{NN}(b)$ \footnote{This expression assumes that
neighborhood is a symmetric concept, such that $a\in \mathrm{NN}(b)$ implies that $b\in
\mathrm{NN}(a)$. This is correct for the definitions of neighborhood based on cutoff
radii and on the Delaney triangulation, but not for the definition based on a
fixed number of neighbors.}. The set of all bonds is called the {\em bond network}.
The idea of bond orientation analysis is to derive scalar metrics from the information
of the bond network (i.e.~the set of bond vectors). The precise definition of the bond
network is therefore crucial.

Other structure metrics are defined in a similar way,  differing only in the geometric interpretation of the bond network, such as ~centro-symmetry metrics
\cite{KelchnerPlimptonHamilton:1998} or Edwards configurational tensors
\cite{PhysicaA01_Edwards} and fcc/hcp-order metrics
\cite{Bargiel:2001:0921-8831:533}, or the number of bonds as the
most simple topological characteristic \cite{Armstrong20093060}.

For a sphere $a$ the set of unit vectors $\mathbf{n}_{ab}$ point from $a$ to the
spheres $b \in \mathrm{NN}(a)$
in the neighborhood of $a$.
Each vector $\mathbf{n}_{ab}$ is
characterized by its angles in spherical coordinates $\theta_{ab}$ and
$\varphi_{ab}$ on the unit sphere.
Following Steinhardt {\em et al.} \cite{PRB83_Steinhardt}, the
local BOO $q_l(a)$ of \emph{weight} $l$ assigned to sphere $a$ is defined as
\begin{equation}\label{eq:definition_q6}
     q_{l}(a)= \sqrt{\frac{4\pi}{2l+1}\sum_{m=-l}^{l}\left\vert \frac{1}{n_a}
\sum_{b\in \mathrm{NN}(a)}
        Y_{lm}\left(\theta_{ab},\varphi_{ab}\right) \right\vert^2},
\end{equation}
where $Y_{lm}$ are spherical harmonics (see e.g.~appendix in
\cite{GrayGubbins}).
This formula can be interpreted as the lowest-order
rotation-invariant (that is, independent of the coordinate system in which
$\theta_{ab}$ and $\varphi_{ab}$ are measured) of
the $l$-th-moment in a multipole expansion of the bond vector distribution
$\rho_\mathrm{bond}(\mathbf{n})$ 
on a unit sphere. Higher-order invariants, often termed $w_l$,
are defined in a similar 
way \cite{PRB83_Steinhardt,wigner1959gruppentheorie}~
\footnote{Although we will not use the global bond order parameter $Q_l$
we define it for completeness as 
$
    Q_{l}= \sqrt{\frac{4\pi}{2l+1}\sum_{m=-l}^{l}\left\vert
\frac{1}{\mathcal{N}} \sum_{k=1}^N \sum_{j=1}^{n_a}
        Y_{lm}\left(\theta_j,\varphi_j\right) \right\vert^2},
$
where $N$ is the number of spherical particles and $\mathcal{N}=\sum_{a=1}^N n(a)$ the
number of all bonds.
This is, the average over all bonds is taken
inside the norm. For disordered systems the sum over the $Y_{lm}$
vanishes as $\mathcal{N}^{-1/2}$, while it remains finite
for common crystalline structures \cite{PRB83_Steinhardt,JChemPhys96_Rintoul}.
}.

The existence of spheres  with values of $q_4$ and $q_6$ 
close to those of an ideal ordered structure (see Tab.~\ref{tab:ql}) 
has been interpreted as evidence of ordered clusters. 
The local structure metrics $q_l$ have been used to identify fcc,
hcp, bcc or icosahedral structures in condensed matter and plasma physics (e.\,g.~in colloidal
particle systems \cite{JChemPhys11_Ni}, random sphere packings \cite{PRE05_Aste,
JChemPhys06_Wouterse} or plasmas \cite{Uspekhi11_Klumov}) by analyzing histograms over the $(q_4,
q_6)$-plane or combinations of similar order parameters \cite{JChemPhys08_Lechner}. 
Frequently, histograms of one order parameter only, namely $q_6$, are used to
qualitatively compare disorder in particulate matter systems
\cite{JChemPhys10_Yiannourakou, EPJB06_Lochmann,
EPJE10_Xu, PRE08_Martin}.
Our previous work\cite{KapferMickelSchroederTurkMecke:2012a} has raised the caution that local configurations can exist
that are clearly non-crystalline but have the same values of $q_6$ as hcp or fcc
environments.
Several authors have defined bond order functions \footnote{
Normalized bond order functions are 
 $q_{lm}(a):=\left(\sum_{i=1}^{n(a)}Y_{lm}\right)/q_l(a)$ for particle
$a$ and the dot-product is $d_{ab}:=\sum_{m=-l}^{l} q_{lm}(a)q_{lm}^*(b)$ of
spheres $a$ and $b$. A particle is defined as member of a solid-like cluster, if the
dot-product with $n_0$ NN exceeds a certain threshold $d_0$.}
closely related to the $q_l$
for the identification of crystalline clusters \cite{PRL10_Schilling,
PRE10_Mokshin, PRL07_Keys, PNAS09-Hernandez-Guzman}.

As a different application from the identification of locally crystalline
domains, it has been proposed to use averages $\langle q_l\rangle $ over all
spheres to quantify the degree of order of a configuration. Averages $\langle
q_6\rangle$ have been analyzed (as function of some control parameter such as
temperature, pressure, strain, or packing fraction) for random sphere packings
\cite{EPJB06_Lochmann}, granular packing experiments \cite{PRE10_Panaitescu},
model fluids \cite{JChemPhys06_deOliveira}, molecular dynamics simulations of water
\cite{PRE07_Yan} or polymer melts \cite{PRL06_Wallace}.
This use of $\langle q_l\rangle$ to quantify the overall degree of order implies
a monotonous relationship between the value of $q_l$ and the
degree of order. In contrast to the identification of individual crystalline
cells as those with $q_l$ the same as for the crystalline reference cell
$q_l^\mathrm{cryst}$, one now
assumes that larger values of $\Delta:=|q_l-q_l^{\mathrm{cryst}}|$ correspond to
``larger'' deviations from the crystalline configuration, even for clearly
acrystalline local configurations with large values of $\Delta$. The validity of
this assumption is difficult to assert, in the absence of an independent
definition of the degree of the ``deviation from crystalline structure''. (Note
also the obvious problem for the case of monodisperse hard spheres, where two
distinct crystal reference states, fcc and hcp, exist which however have
different values of $q_l$.)
Nevertheless, $q_6$ has been used to quantify order in disordered packings,
under the assumption that higher values of $q_6$ correspond to higher degree of order
\cite{PRE02_Kansal}.
Unless the system
represents a small perturbation of one specific crystalline state, this use of
$q_6$ is, in our opinion, not justified. $q_6$ is not a suitable order metric to
compare the degree of order of disordered configurations that are far away from
a crystalline reference state. We use the term \emph{structure metric} to
emphasize that a priori $q_l$ does not quantify order in disordered systems. 

\begin{table}[t]
 \begin{tabular}{l|c|c|c|c|c|c}
         & \multicolumn{2}{|c|}{bcc}          & fcc       & hcp     & icosahe-
& simple cubic \\
&\multicolumn{2}{|c|}{$Im\bar{3}m$} & $Fm\bar{3}m$ & $P6_3/mmc$ & dral & $Pm\bar{3}m$ \\
         & $n=8$           & $n=14$     & $n=12$  & $n=12$& $n=12$
& $n_a=6$   \\\hline
 $q_2$   & $0$               & $0$          & $0$       & $0$     & $0$
& $0$       \\
 $q_3$   & $0$               & $0$          & $0$       & $0.076$ & $0$
& $0$       \\
 $q_4$   & $0.509$           & $0.036$      & $0.190$   & $0.097$ & $0$
& $0.764$   \\
 $q_5$   & $0$               & $0$          & $0$       & $0.252$ & $0$
& $0$       \\
 $q_6$   & $0.629$           & $0.511$      & $0.575$   & $0.484$ & $0.663$
& $0.354$   \\
 $q_7$   & $0$               & $0$          & $0$       & $0.311$ & $0$
& $0$       \\
 $q_8$   & $0.213$           & $0.429$      & $0.404$   & $0.317$ & $0$
& $0.718$   \\
 $q_9$   & $0$               & $0$          & $0$       & $0.138$ & $0$
& $0$       \\
 $q_{10}$& $0.650$           & $0.195$      & $0.013$   & $0.010$ & $0.363$
& $0.411$   \\
 $q_{11}$& $0$               & $0$          & $0$       & $0.123$ & $0$
& $0$       \\
 $q_{12}$& $0.415$           & $0.405$      & $0.600$   & $0.565$ & $0.585$
& $0.696$   \\
 \end{tabular}
\caption{\label{tab:ql} Values of $q_l$ in perfectly symmetric configurations. 
For these highly symmetric cases (fcc, hcp, icosahedron,sc), the definitions of neighborhood discussed in
this article all yield the same crystallographic neighbors, and hence values of
$q_l$ (assuming infinite precision for the point coordinates such that the
Delaunay diagram has edges to {\em all} nearest crystallographic neighbors).
Spheres in bcc configuration have 8 nearest neighbors at distance $\sigma$,
where $\sigma$ is the particle diameter, and 6 second nearest neighbors
at distance $\sqrt{2}\sigma$ and have 14 Delaunay neighbors. }
\end{table}

We here demonstrate a further aspect, distinct to those described above,  that
should be taken into account when interpreting $q_l$ data for disordered
systems, namely a very significant dependence of the $q_l$ values on details of
the definition of the bond network: changes of the NN definition do not only
affect the absolute values (which are of great importance, as the comparison to
the crystalline reference values is in terms of these absolute values) but they
can also affect functional trends. This observation highlights the problem in
the interpretation of anomalies of the BOO parameters (that is, 
local extrema as function of some thermodynamic parameter)
as being
connected to thermodynamic anomalies \cite{JChemPhys06_deOliveira, PRE07_Yan}; 
see also the discussion of the anomalies of water \cite{Natur01_Errington} in
terms of a parameter similar to the BOO parameters.
Rather than being a mere inconvenience, the dependence on the details of the
bond network definition is of direct relevance to the physical interpretation.

\section*{Ambiguity of the neighborhood definition and its 
effect on $q_l$\label{sec:ambiguity}}
\begin{figure}[t]
\begin{center}
\begin{tabular}{ll}
 a) & b) \\
 \includegraphics[bb=460 185 1240 820, clip,
width=4cm]{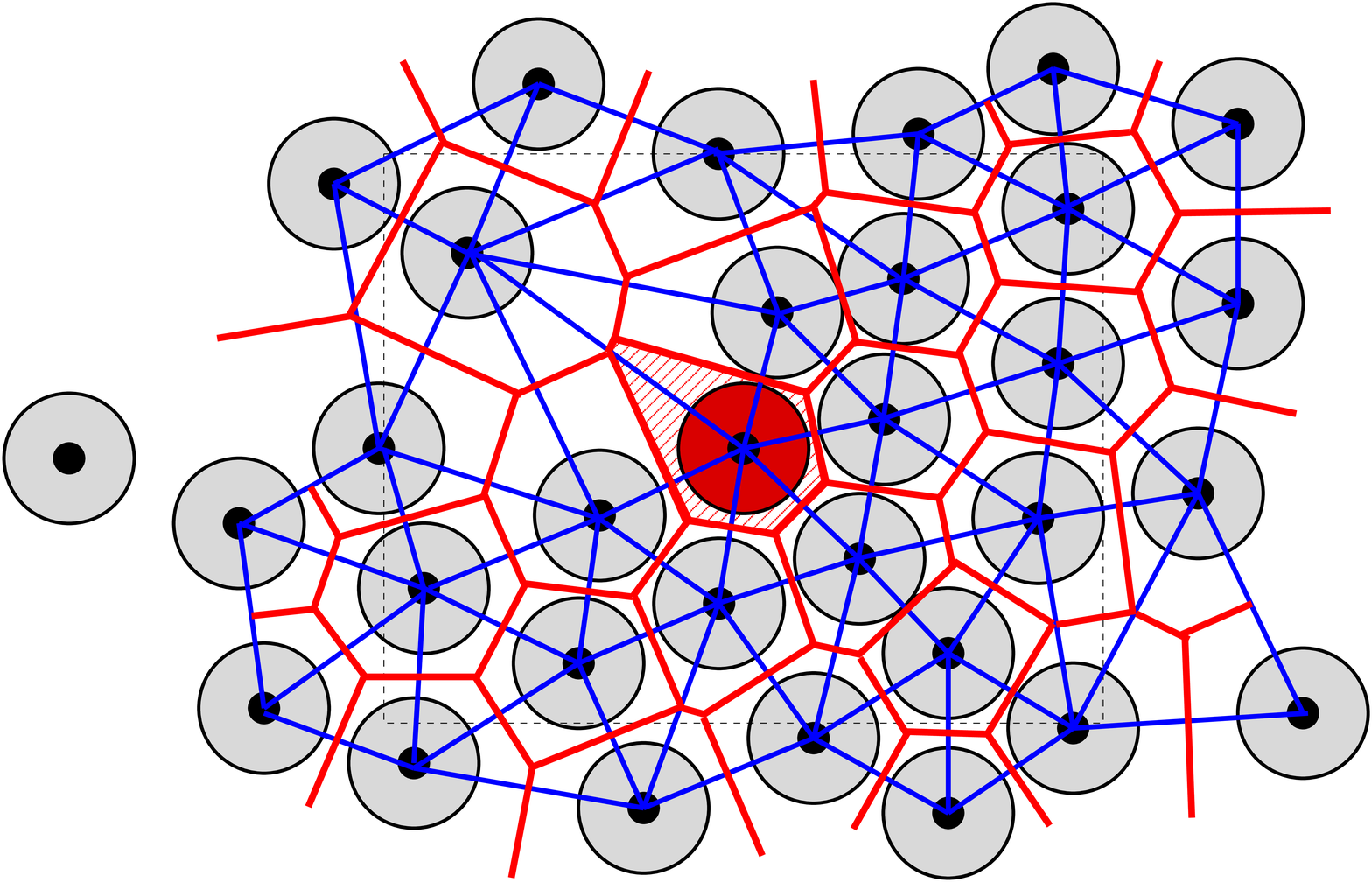} & 
 \includegraphics[bb=460 185 1240 820, clip,
width=4cm]{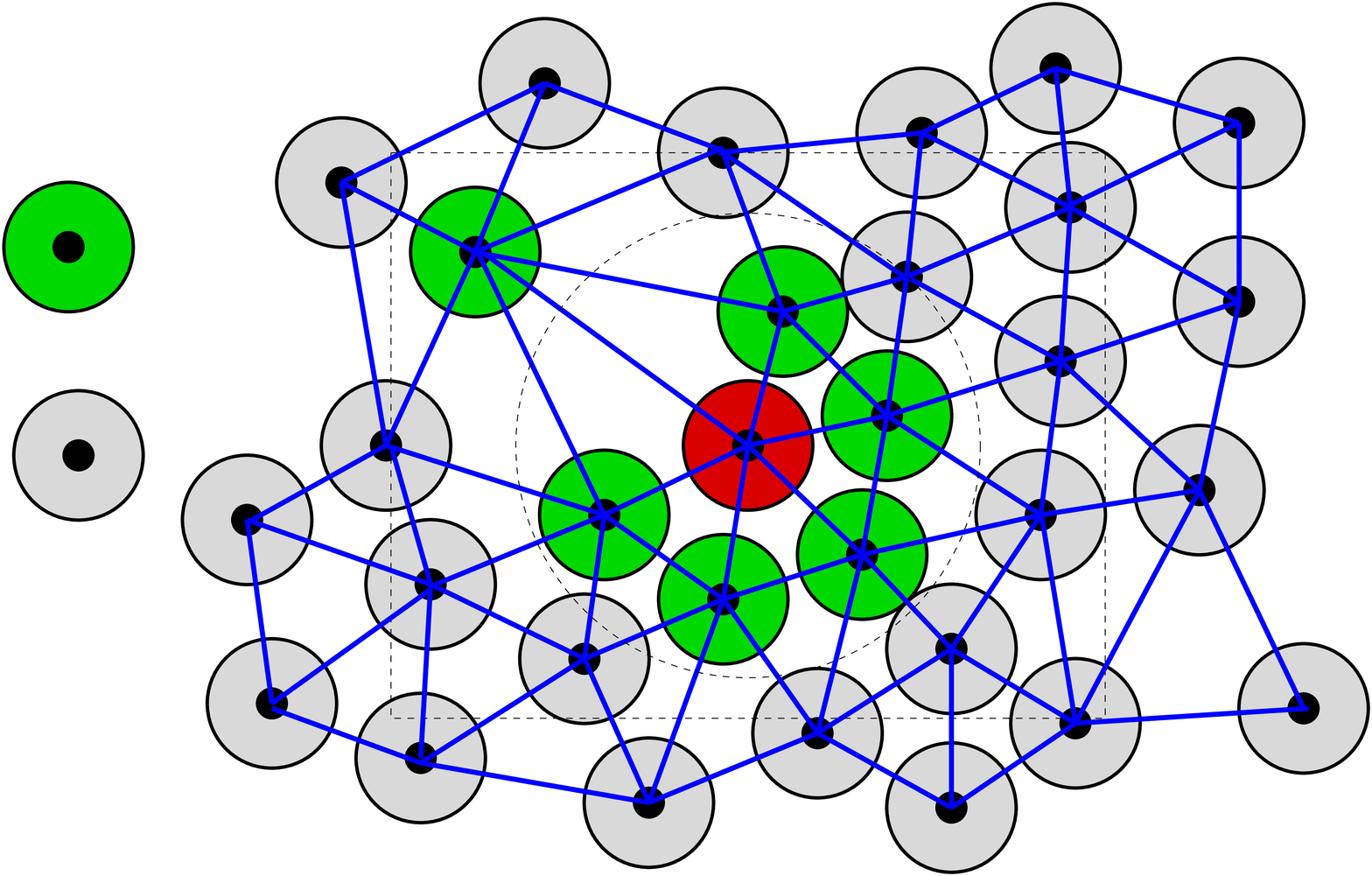} \\
c) & d) \\
 \includegraphics[bb=460 185 1240 820, clip,
width=4cm]{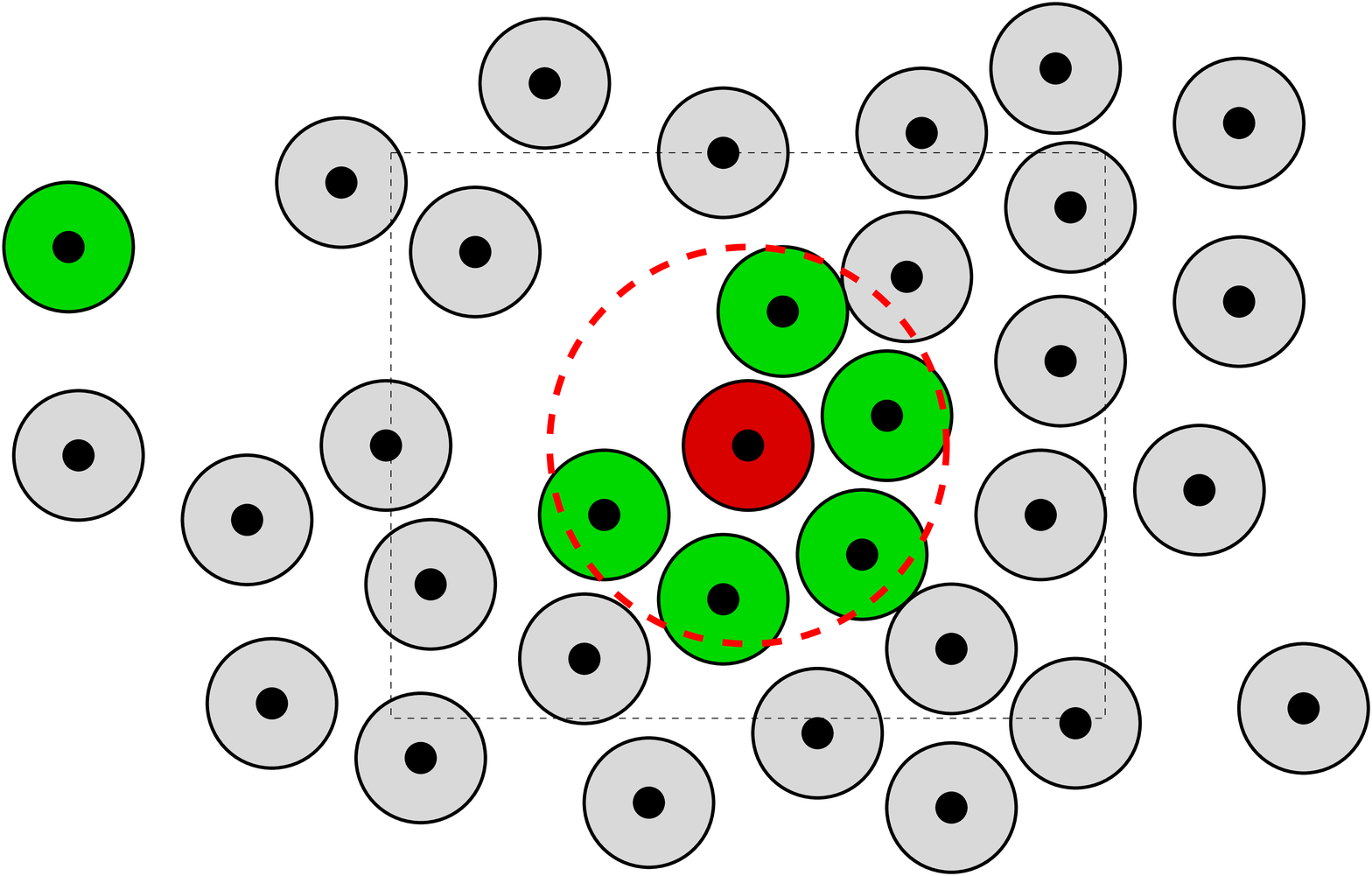} & 
 \includegraphics[bb=460 185 1240 820, clip,
width=4cm]{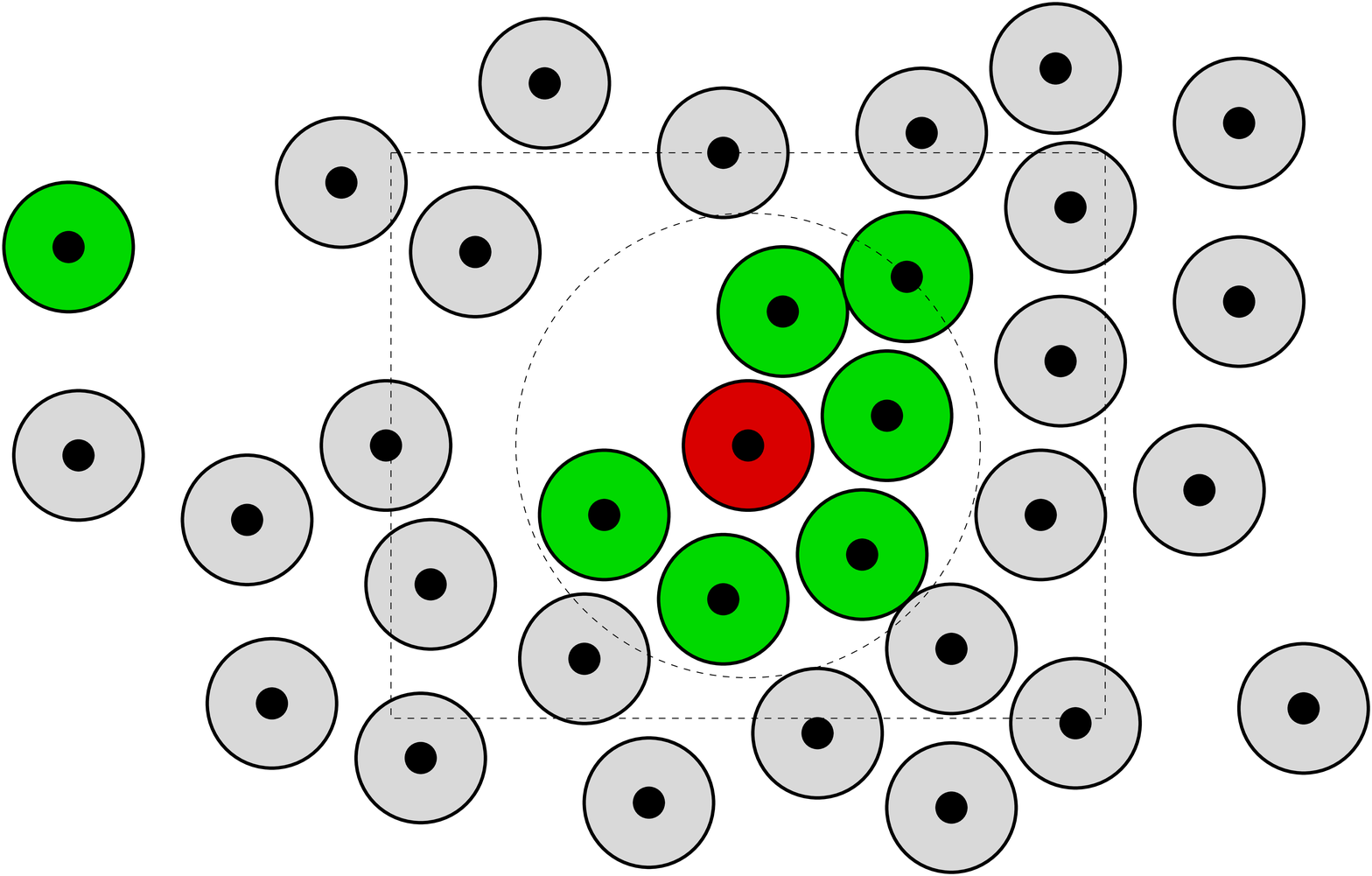} \\
\end{tabular}
\end{center}
\caption{(color online)
\label{fig:neighbors}
Widely used NN definitions: 
a) Voronoi diagram (red) and its dual, the Delaunay graph (blue)
b) Delaunay definition of nearest neighbors (NN): the Delaunay
neighbors of the red sphere are highlighted in green.
c) NN definition with cutoff radius $r_\mathrm{c}$
d) $n_{\mathrm{f}}$ closest NN, here $n_{\mathrm{f}}=6$
}
\end{figure}

The choice of a set of nearest neighbors
-- at the heart of bond orientation analysis -- is not unique (see
Fig.~\ref{fig:neighbors}).
Steinhardt \emph{et al.} proposed to use ``some suitable set'' of bonds for
the computation of $q_l$; they used a definition based
on a cutoff radius of $1.2\sigma$, where $\sigma$ is the particle diameter
\cite{PRB83_Steinhardt}. That is, each sphere that is closer to a given sphere $a$
than a cutoff radius $r_c$ is assigned as a NN of sphere $a$.
Neighborhood definitions based on cutoff radii are widely used,
 e.g.~with cutoff radii $1.2\sigma$ and $1.4\sigma$
\cite{JChemPhys09_Mokshin,JChemPhys09_Odriozola,PRE08_Martin,PRE07_Duff,
PRL07_Keys} or with the value of the cutoff radius determined by the
the first minimum of the two-point correlation function $g(r)$ \cite{PRE10_Kurita,
JChemPhys09_Calvo,PNAS09-Hernandez-Guzman, PRE08_Abraham,PRE07_Wang}.

Alternatively, the Delaunay graph of the particle centers
\cite{BarberDobkinHuhdanpaa:1996}~
\footnote{The definition of NN via the Delaunay graph
is equivalent to the definition via Voronoi neighbors: spheres
share a Delaunay edge, whenever their respective Voronoi cells have a shared facet
(regardless of the area of the Voronoi facet).}
is used to define NN
\cite{EPJE10_Xu, PRE10_Panaitescu, JChemPhys06_Wouterse,JChemPhys06_Kumar, EPJB06_Lochmann}.
In this parameter-free
method, every sphere which is connected to a sphere $a$ by a Delaunay edge
is considered a NN of $a$.
A rarely used definition is to assign a fixed number $n_{\rm f}$ of NN to each
particle $n(a)=n_{\rm f}$ \cite{JChemPhys06_deOliveira,PRE07_Yan}. In three
dimensions, the $n_{\rm f}=12$ other spheres closest to the central sphere are chosen
as neighbors.
The difference between these definitions is  illustrated in Fig.~\ref{fig:neighbors}.
Note that while the definitions via cutoff radius and via the Delaunay graph are
symmetric, i.\,e.~$b\in \mathrm{NN}(a)\Leftrightarrow a\in \mathrm{NN}(b)$, the definition of
neighborhood as the nearest $n_{\rm f}$ spheres is not, see Fig.~\ref{fig:neighbors}
(d). The definitions of NN discussed so far will be called 
\emph{bond network neighborhoods} in the following; in this picture, each nearest neighbor
is equivalent to the other neighbors.  By contrast, we use the term \emph{morphometric
neighborhood} if the neighborhood relation is additionally weighted with
geometrical features.

A principal weakness of structure metrics based on bond network neighborhoods is their lack
of robustness: Small changes of particle positions can delete or add entries in the
set of neighbors. This discontinuity w.\,r.\,t.~the particle positions is inherited
by the structure metrics defined via bond network neighborhoods.
Small changes in the particle coordinates can lead to large changes in the structure
metrics, which is undesirable.

We demonstrate the very strong effect of the NN definition on the BOO parameter
$q_6$ by the example of a super-cooled fluid. Using non-equilibrium
molecular dynamics (MD) simulations \cite{arxiv10:Nogawa, JSTAT10_Kapfer}~
\footnote{Event driven MD simulations to explore the super-cooled regime use the Matsumoto
algorithm from Ref.~\cite{arxiv10:Nogawa}. In this algorithm, spheres are expanded until they touch the closest Voronoi facet or
until they reach the final radius. This creates a transient polydisperse
ensemble, which is relaxed by thermal motion, followed by an expansion step. This
procedure is iterated until a monodisperse HS system at predefined packing fraction is obtained.},
super-cooled configurations are generated that represent entirely disordered
states with densities larger than the 
fluid-crystal coexistence density of hard spheres (HS)
of $\phi\approx0.494$
\cite{Nature97_Woodcock}.

\begin{figure}[t]
 \includegraphics[
width=1\columnwidth]{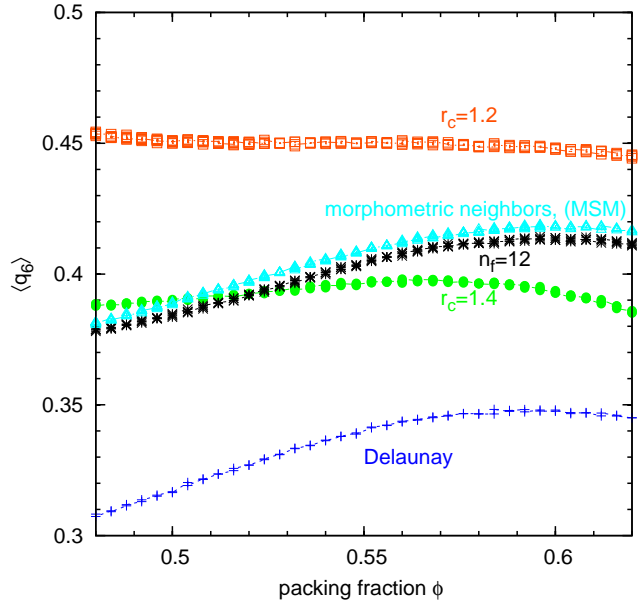}
\caption{\label{fig:zoomin_super-cooled_liquid-averages-q6}
(color online)
Average local bond order parameter $\langle q_6\rangle $ in the super-cooled
HS fluid with
several definitions of the nearest neighbors: orange squares:
$r_\mathrm{c}=1.2\sigma$, green bullets: $r_\mathrm{c}=1.4\sigma$ blue crosses:
Delaunay definition and black stars: $n_{\rm f}=12$. The turquoise triangles represent data for the Minkowski structure metrics (MSM) $\langle q_6'\rangle$ defined in Eq.~\eqref{eq:definition_q6prime}.
}
\end{figure}

Figure \ref{fig:zoomin_super-cooled_liquid-averages-q6} shows the average local
BOO $\langle q_6\rangle$ for four different choices of bond network neighborhood definition.
To distinguish between the different definitions of neighborhood discussed above,
we use the symbols $q_6^{r_\mathrm{c}}$, $q_6^\mathrm{D}$ and $q_6^{n_{\rm f}}$.
First, the absolute values of $q_6^{r_c=1.2\sigma}$, $q_6^{r_c=1.4\sigma}$,
$q_6^{n_{\rm f}=12}$ and
$q_6^{\rm D}$ differ significantly, which is important when comparing these values
to that of a specific crystalline phase
such as fcc.
Second, and of greater concern for the use of $q_6$ as a structure metric, the
behavior of $q_6^{r_c=1.2\sigma}$, $q_6^{r_c=1.4\sigma}$, $q_6^{n_{\rm f}=12}$ and $q_6^{\rm D}$ is
qualitatively different as a function of the packing fraction $\phi$. For
example $\langle q_6^{r_c=1.2\sigma}\rangle(\phi)$ shows a slight negative trend
without pronounced extrema, whilst $\langle q_6^{r_c=1.4\sigma}\rangle(\phi)$ increases for
$\phi < 0.56$ and decreases above. $\langle q_6^{n_{\rm f}=12}\rangle(\phi)$
and $\langle q_6^{\rm D}\rangle(\phi)$ show a maximum at slightly different
positions with a significantly different absolute value. 
Each of these trends is specific to the neighborhood definition.
These discrepancies raise a caution flag about the use of $q_6$ as a local structure
metric in disordered systems. This is in accordance with several reported difficulties in the application
of $q_6$ in ordered and disordered systems \cite{PRE02_Kansal, JChemPhys96_tenWolde, EPL98_Troadec}.
 The choice of the NN definition has a
dominant effect on the values and on the functional trend of $\langle
q_6\rangle(\phi)$ that conceals the behavior due to genuine structural changes
induced by the physics of the system.
Results for $q_6$ obtained by different studies are not only difficult to
compare quantitatively, but also the qualitative behavior may be misleading.

\begin{figure}[t]
 \includegraphics[width=1\columnwidth]{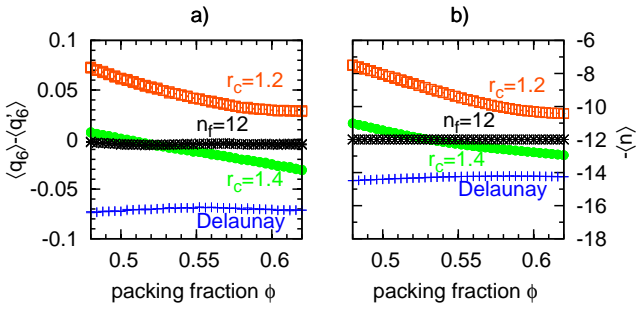}
\caption{
\label{fig:correlation-q6-volume}
(color online)
(a) Average number of nearest neighbors identified by the different
definitions of neighborhood, for the same data as shown in
Fig.~\ref{fig:zoomin_super-cooled_liquid-averages-q6}.
Difference $\langle q_6^{X} \rangle - \langle q_6'\rangle$ between $q_6$
values for different definitions of the bond network neighborhood
$X = \{{r_c=1.2\sigma}, {r_c=1.4\sigma}, {\rm D}, n_{\rm f}=12\}$.
(b)~Comparison of the functional
trend of these data to $-\langle n\rangle(\phi)$ demonstrates the strong
negative correlation of the value of $q_6$ with the number of NN
spheres $n$ identified by the specific neighborhood definition.
}
\end{figure}

The behavior of $\langle q_6\rangle$ %
can be rationalized by considering the average number of nearest neighbor spheres $\langle n\rangle(\phi)$ identified by the
different neighborhood definitions.

Figure \ref{fig:correlation-q6-volume} (a) shows $\langle q_6 \rangle-\langle q_6' \rangle$ as function of $\phi$.
$q_6^\prime$ is a structure metric based on morphometric neighborhood, which is discussed in detail in the following section.
Figure \ref{fig:correlation-q6-volume} (b) shows $-\langle n\rangle$. These data demonstrate a very close correlation between $\langle q_6 \rangle-\langle q_6' \rangle$ and $-\langle n\rangle$, valid for all neighborhood definitions. This result asserts that $\langle q_6'\rangle$
captures physical structure properties, while various variants of $\langle q_6\rangle$ are predominantly indicative of the
typical number of NN spheres $\langle n\rangle$ identified by the respective NN definitions.

\begin{figure}[t]
 \includegraphics[width=0.65\columnwidth]{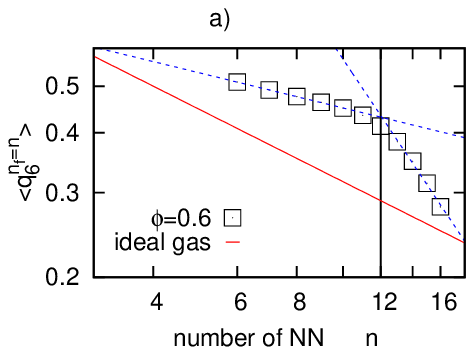}
 \includegraphics[width=0.25\columnwidth, bb=0 37 122 284,clip]{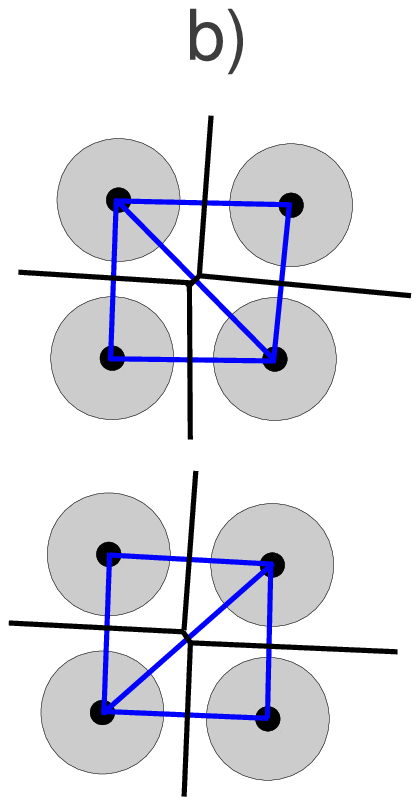}
\caption{\label{fig:q6_vs_zNN}
(color online)
(a) Mean $\langle q_6^{n_{\rm f}=n}\rangle $ as a function of the fixed number
$n$ of neighbors assigned to each sphere.
The squares are data of a super-cooled fluid with $\phi=0.6$ and the red solid
line of the ideal gas ($\langle q_6^{n_{\rm f}=n}\rangle  \propto n^{-1/2}$, see
Ref.~\cite{JChemPhys96_Rintoul}).
 The dotted lines are fits for the first coordination shell for $n<12$
 and first and second shell $n>12$. The first shell exponent is $-0.24$ and
the
 second shell exponent is $-1.48$.
(b) Illustration for the discontinuity of the topology of the Voronoi diagram as
function of center point coordinates: An infinitesimal particle displacement can
destroy or create Voronoi cell facets (and hence bonds in the neighborhood
definition based on the Delaunay graph).}
\end{figure}

Figure~\ref{fig:q6_vs_zNN} further corroborates
this observation by the analysis
of $\langle q_6^{n_{\rm f}=n}\rangle$ as a function of $n$ for
the  super-cooled hard sphere fluid at $\phi=0.600$. The average $\langle
q_6^{n_{\rm f}=n}\rangle(n)$ systematically
decreases with higher prescribed numbers $n_{\rm f}$ of NN. This effect is further
amplified for large $n_{\rm f}>12$, when spheres in the second coordination
shell are also identified as neighbors.
The stronger decrease in $q_6$ when encountering the second coordination shell
also explains
why $\langle q_6^{\rm D}\rangle$ generally has lower values compared to the other
neighborhood definitions, since the typical number of Delaunay neighbors is higher than for the other neighborhood definitions, $\langle n_a^{\rm D}\rangle\approx 14$.

\section*{Minkowski structure metric by Voronoi-cell weighting \label{sec:q6prime} }
This section introduces the Minkowski structure metrics $q_l'$ that were already alluded to above.
The Minkowski structure metrics (MSM) are obtained by an
adaption of the conventional BOO parameters. The MSM differ from the conventional
$q_l$, Eq.~\eqref{eq:definition_q6}, by the fact that the contribution of
each neighbor to the structure metric is weighted by an associated relative area factor $A(f)/A$.
In this factor, $A(f)$ is the surface area of the Voronoi cell facet $f$
separating the two neighboring spheres that correspond to a given bond, and $A=\sum_{f\in\mathcal{F}(a)}A(f)$
is the total surface area of the Voronoi cell boundary $\mathcal{F}(a)$ of sphere $a$.
This simple change leads to
robust, continuous and parameter-free structure metrics $q_l'$ that avoid the
shortcomings of the conventional $q_l$ discussed above.

We define
\begin{equation}\label{eq:definition_q6prime}
     {q}_{l}'(a)= \sqrt{\frac{4\pi}{2l+1}\sum_{m=-l}^{l}\left\vert \sum_{f\in
\mathcal{F}(a)}
       \frac{A(f)}{A} Y_{lm}\left(\theta_f,\varphi_f\right) \right\vert^2},
\end{equation}
where $\theta_{f}$ and $\varphi_{f}$
are the spherical angles of the outer normal vector $\mathbf{n}_f$ of facet $f$.
Note that the direction of this vector coincides with the bond vector that is
used in conventional bond orientation analysis (see Fig.~\ref{fig:neighbors}).

Because of the weighting of each bond by its corresponding Voronoi facet area
$A(f)/A$, these newly constructed structure metrics $q_l'$ are continuous functions
of the spheres' center point coordinates, and hence robust. Furthermore, this
geometrical neighborhood is symmetric and parameter-free.

The definition of
$q_l'$ results naturally from a multipole expansion in spherical harmonics  of
the Voronoi cell surface normal distribution function 
\begin{equation}
\rho(\mathbf{n})=\frac 1A\cdot\sum_{f\in\mathcal{F}}\delta\bigl(\mathbf{n}
(f)-\mathbf{n}\bigr)A(f)
\end{equation}
on the unit sphere: $\rho(\mathbf{n})=\rho(\theta,\varphi)=\sum_{l=0}^\infty\sum_{m=-l}^l q_{lm}'Y_{lm}(\theta,\varphi)$, where
$q_{lm}'$ evaluates to $\sum_{f\in\mathcal{F}(a)} ( A(f)/A ) Y_{lm}^*(\theta_f,\varphi_f)$;
the star denoting complex conjugation.

By contrast, the $l$th-moment of the distribution $\rho(\mathbf{n})$ in Cartesian coordinates is
\begin{equation}\label{eq:NMT}
 W_1^{0,l}:= \sum_{f\in\mathcal{F}} \underset{l\text{
times}}{\underbrace{\mathbf{n}(f)\otimes\ldots\otimes\mathbf{n}(f)}} A(f),
\end{equation}
where $\otimes$ denotes the tensor product. The moment tensors
$W_1^{0,l}$ are special types of Minkowski tensors \cite{JSTAT10_Kapfer,
AdvMatt11_SchroederTurk}. These versatile shape metrics have been studied in the field of integral geometry \cite{SchneiderWeil} and
successfully applied to analyze structure in jammed bead packs \cite{ EPL10_SchroederTurk, PRE12_Kapfer},
bi-phasic assemblies \cite{DoiOhta:1991,Langmuir11_SchroederTurk}, foams \cite{Evans:2012} and other cellular structures \cite{AdvMatt11_SchroederTurk, PRE12_Evans}.
There is a one-to-one 
correspondence between this class of Minkowski tensors and the multipole
expansion of the surface normal vector distribution $\rho(\mathbf{n})$
of a convex Voronoi polytope $\mathcal{F}(a)$
\cite{KapferMickelSchroederTurkMecke:2012b, AdvancesInPhysics78_Jerphagnon}.

For ideal crystals where all Voronoi facets have equal size, the values of the BOO $q_l$ and of
the MSM $q_l'$ are the same; these symmetries are fcc, hcp, the icosahedron and sc (simple cubic).
In the case of bcc, where Voronoi cells have in total 14 facets, of which 8 correspond to closest neighbors
and 6 to neighbors in the second shell, $q_l$ differ from $q_l'$ (see also Table~\ref{tab:ql}).

The construction of the weighted $q_l'$ has no adjustable parameters.
However, the choice of the Voronoi diagram as the partition that
defines local neighborhood and that is used for the definition of
$q_l'$ may be viewed as arbitrary. Its use can be justified as follows: First, the use of any partition of space into cells associated with the beads for the neighborhood definition guarantees symmetric neighborhoods, $\left(a\in \mathrm{NN}(b)\right) \Leftrightarrow \left(b\in \mathrm{NN}(a)\right)$. Second, the use of the Voronoi diagram ensures that the following minimal requirements are met: (a) convex cells, (b) invariance under exchange of spheres decorating the seed points and (c) the possibility to reconstruct the seed point coordinates uniquely from the facet information \cite{Lautensack:2007}. The authors are unaware of an alternative to the Voronoi diagram that fulfills these requirements.

\section*{\label{sec:interpretation-different-weights}
Geometric interpretation of the Minkowski structure metrics, in particular of $q_2'$}

For the use of both BOO parameters and MSM, an important issue is the choice
of the weights $l$ that are considered.  Many studies restrict themselves
to only $q_6$, possibly supplemented by $q_4$ and the associated higher-order
invariants $w_4$ and $w_6$.  
This is likely to be motivated by $q_6$ being the apparent generalization
of the two-dimensional hexatic order parameter $\psi_6$. The relation between
the $l=6$ structure metrics and ordering, however, is not as direct
in 3D as it is in 2D:  $q_6$ is maximized by icosahedral bond order,
which is incompatible with translational order.  The perception that large values
of certain structure metrics, in particular $q_6$, are intrinsically connected
with crystallization is therefore deceiving, and it is useful to discuss
the relevance of the individual weights to physical problems.

In all cases, $q_0'$ is trivially $1$ while $q_1'$ trivially vanishes, due to the so-called {\em envelope theorems} of Mueller \cite{Mueller:1953}  (note, this does not apply to $q_1$). Thus, the first weight that captures pertinent information about a disordered system is $l=2$; for hcp and fcc crystals $q_2'$ vanishes.
The invariants $q_3'$ and $q_5'$ (and odd weights in general) vanish in 
configurations symmetric under inversion, but capture deviations from
this symmetry (see tab.~\ref{tab:ql}).
Hence they might be robust candidates for defect detection like centro-symmetry
metrics \cite{KelchnerPlimptonHamilton:1998} or
to separate hcp from fcc, since the hcp Voronoi cell is not inversion symmetric
(see tab.~\ref{tab:ql}), while
fcc is inversion symmetric ($m\bar{3}m$) with respect to the sphere centers.
Including Steinhardt {\it et al.}'s original paper \cite{PRB83_Steinhardt} we are not aware
of any applications of odd weights $l$. 
The lowest weight to discriminate a sphere from a cube is $l=4$ and thus plays
an important role in ordered materials.
The cubic-symmetry fcc, bcc, and simple cubic lattices all have
non-vanishing $q_4'$ values (for the conventional BOO parameters though,
great care is needed for the bond definition,
as different sets of NN for bcc reveals a dramatic change on conventional $q_4$).
$q_6$ is the first non-vanishing weight for icosahedral symmetry (and maximum for
the icosahedron).  Note that the $q_6$ values for fcc can be matched by deformed
icosahedral bonds.

\begin{figure}
 \includegraphics[width=0.8\columnwidth]{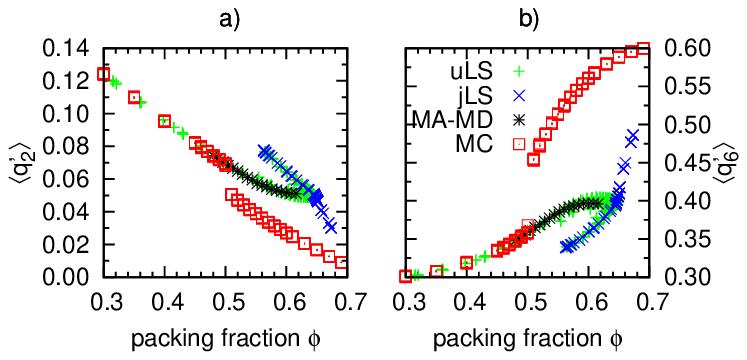}
 \includegraphics[width=0.8\columnwidth]{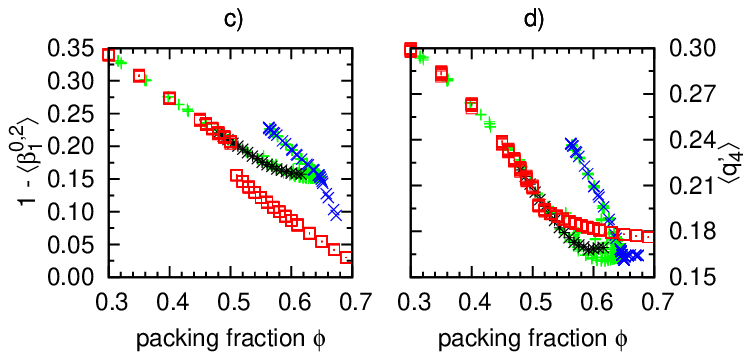}
\caption{
(color online)
Minkowski structure metrics $q_2'$, $q_4'$ and $q_6'$ for
equilibrium hard spheres (Monte Carlo, MC) \cite{JSTAT10_Kapfer}
simulations, jammed Lubachevsky-Stillinger (jLS) \cite{EPL10_SchroederTurk},
non-equilibrium unjammed Lubachevsky-Stillinger (uLS)
\cite{SkogeDonevStillingerTorquato:2006}
and non-equilibrium Matsumoto algorithm (MA-MD) simulations
\cite{arxiv10:Nogawa} (see text).
$\beta_1^{0,2}$ is the anisotropy index, i.e.~the ratio of the
smallest and the largest eigenvalue of the Minkowski tensor
$W_1^{0,2}$; see Eq.~(\ref{eq:NMT}) and Ref.~\cite{EPL10_SchroederTurk}.
\label{fig:StructureDiagram}}
\end{figure}

While in ordered states, the $q_l$ are easily interpreted, in disordered states 
the lack of a well-defined reference state renders the interpretation more difficult.
Fig.~\ref{fig:StructureDiagram} shows $\langle q_2'\rangle$, $\langle
q_4'\rangle$ and $\langle q_6'\rangle$ of hard-sphere
systems in a wide range of packing fractions. The plot includes data from Monte Carlo
simulations of the thermal equilibrium fluid/solid \cite{JSTAT10_Kapfer} (MC),
from fully disordered and partially crystalline jammed Lubachevsky-Stillinger (jLS)
\cite{JStatPhys90_Lubachevsky, EPL10_SchroederTurk}, and also from unjammed
non-equilibrium simulations (uLS) from LS simulations before jamming \footnote{
In the LS algorithm \cite{SkogeDonevStillingerTorquato:2006} spheres are continuously expanded with event-driven MD until
the pressure exceeds a jamming threshold (jLS). The unjammed LS simulations (uLS) used here are stopped at predefined packing fractions.
} and the data from
Fig.~\ref{fig:zoomin_super-cooled_liquid-averages-q6} (MA-MD) \cite{arxiv10:Nogawa}.

Empirically, we find that disordered  cells virtually always have finite $q_2'$
values; for order (cubic-symmetry or close packed), $q_2'$ vanishes.
Therefore, distributions of $q_2'$ in a partially ordered system are bimodal,
which is convenient for the separation of both phases.
Conversely, if the abundance of small values $q_2' \approx 0$ in a sample
vanishes, one can conclude that it is fully disordered.
The information contained in the lowest weight $q_2'$ is also captured in
the anisotropy index $\beta_1^{0,2}$ derived from Minkowski tensors\footnote{
The anisotropy index $\beta_1^{0,2}$ is the ratio of
eigenvalues  of $W_1^{0,2}$ (see Eq.~\eqref{eq:NMT}):
$\beta_1^{0,2}=\xi_{\mathrm{min}}/\xi_{\mathrm{max}}$, where
$\xi_{\mathrm{min}}\leq\xi_\mathrm{mid}\leq\xi_{\mathrm{max}}$.
$\beta_1^{0,2}=1$ indicates isotropy, lower values of
$\beta_1^{0,2}$ indicate anisotropy \cite{EPL10_SchroederTurk}.}, see the comparison of $\langle q_2'\rangle$ and $1-\langle \beta_1^{0,2}\rangle$ in Fig.~\ref{fig:StructureDiagram}.

The observation that $q_2'$ vanishes for ordered configurations corresponds
to the fact
that $\beta_1^{0,2}=1$, and $q_2'>0$ corresponds to $\beta_1^{0,2}<1$
(cf.~Refs.~\cite{EPL10_SchroederTurk, KapferMickelSchroederTurkMecke:2012a, JSTAT10_Kapfer}).

Both structure metrics, $q_2'$ and $\beta_1^{0,2}$,
capture well the different features in local structure of hard-sphere systems
(Fig.~\ref{fig:StructureDiagram}, panels (a) and (c)). The thermodynamic phase transition from
the fluid to the solid (fcc) phase at packing fractions around $\phi\approx
0.49$ is clearly visible. Furthermore, jammed sphere packs are well
distinguished from the equilibrium configurations.
Starting from the equilibrium and avoiding crystallization,
the non-equilibrium MA-MD protocol continues the fluid branch into a
super-cooled fluid regime.  The uLS protocol generates further
non-equilibrium states with larger $q_2'$, up to jammed configurations.
In both diagrams
(a) and (c), the non-equilibrium fluid states are found above the linear
extrapolation of the equilibrium
fluid branch, while the ordered phase is below.
The diagram (b), showing $\langle q_6'\rangle$, reproduces
(though ``upside down'')
quite well the qualitative features obtained from $\langle q_2'\rangle$ or
$\langle 1-\beta_1^{0,2}\rangle$.
The agreement of these two plots, however, is coincidental.
While the separation of the fluid and solid branches in the $q_2'$
diagram is due to the fact that only ordered clusters have vanishing
$q_2'$, there is a large number of possible disordered clusters that
have $q_6'\approx q_6^\mathrm{fcc}$, in particular, perturbed
icosahedral bond arrangements.  These are, however, not present
in the data in large numbers and thus can be neglected
\cite{KapferMickelSchroederTurkMecke:2012a, PRL07_Anikeenko}.
If they occurred in significant abundance in the systems, an
increase of $\langle q_6'\rangle$ would be the consequence.
Values of $q_6'$ close to $q_6^\mathrm{hcp}$ do, however, occur
even in disordered systems \cite{KapferMickelSchroederTurkMecke:2012a}.
Thus deviations from $q_2'=0$ arguably are a better criterion
for disorder than deviations from $q_6^\mathrm{fcc}$.

Since both fcc and hcp have $q_2'=0$, they cannot be discerned using $q_2'$
alone. The dense ($\phi>0.649$) jLS packings, for example, consist of a
significant fraction of hcp and fcc clusters on a disordered background.
Increasing packing fraction reduces the amount of disordered configurations,
and proportionally, their weight in the $\langle q_l'\rangle$ averages.
Consequently, the $q_2'$ curves tend towards $q_2'=0$ as the ordered
clusters take over a larger amount of the system, while
the terminus of the $q_6'$ curves reflects an average of $q_6^{\rm fcc}$
and $q_6^{\rm hcp}$, weighted with the relative fraction of fcc and hcp
domains.

A separation of all the regimes can not be seen in the $q_4'$ plot (d), since
$q_4'$ takes for crystalline (fcc and hcp)
phases fixed values which are lying on a strong random background from the
disordered parts of the system.

\section*{Conclusion}
This article has clearly demonstrated that the conventional bond-orientational order parameters $q_l$, defined via nearest neighbor bonds, Eq.~\eqref{eq:definition_q6}, are very strongly affected by the choice of neighborhood definition (cf.~Fig.~\ref{fig:zoomin_super-cooled_liquid-averages-q6}); this sensitivity is observed both in the qualitative trend and in absolute values. It was shown that for disordered systems without crystallization, $q_6$ strongly correlates to the average number of nearest neighbors. This effect overshadows the actual structural changes induced by the physics of the system 
(cf.~Fig.\ref{fig:correlation-q6-volume}). This dependence is a major drawback that needs to be taken into account when using $q_l$ for the analysis of particulate matter, especially when comparing $q_l$ values across different studies. 

We have proposed a unique, well-defined and robust structure metric $q_l'$, Eq.~\eqref{eq:definition_q6prime},
that avoids the ambiguities that come with bond network neighborhoods.  Robustness of the
structure metric is achieved by quantifying the geometry of the Voronoi tessellation.
The MSM share the same mathematical form with the conventional bond-orientational order
parameters, but the ``bonds'' are weighted with the associated Voronoi facet area.
This guarantees, in particular, that the new Minkowski structure metrics are continuous
as a function of the sphere coordinates.
For hcp, fcc and simple cubic lattices, this definition reproduces
the values of the conventional $q_l$  (cf.~Tab.~\ref{tab:ql}).
For super-cooled hard-sphere fluids, the MSM $q_6'$ is very similar to
the conventional $q_6$ with the (rarely used) $n_{\rm f}=12$ neighborhood definition, see Fig.~\ref{fig:zoomin_super-cooled_liquid-averages-q6}.

The morphometric neighborhood has previously been characterized using
Minkowski tensors \cite{EPL10_SchroederTurk,KapferMickelSchroederTurkMecke:2012a,JSTAT10_Kapfer},
which measure the distribution of normal vectors of the Voronoi cells.
The Minkowski structure metrics presented here can be interpreted as the
rotational invariants of a multipole expansion of the same distribution
of normal vectors; indeed, the approaches of higher-rank Minkowski tensors
and Minkowski structure metrics turn out to be mathematically equivalent ways
to cure the shortcomings of bond-orientational order parameters.
There are further possibilities to address this problem by introducing
weighting factors, see for example Ref.~\cite{JChemPhys96_tenWolde}. Note however that these
approaches need adjustable parameters.
The caution for the use of $q_6$ as a sole determinant of local crystallinity
expressed in Ref.~\cite{KapferMickelSchroederTurkMecke:2012a}, however, is independent of
the issues addressed by this paper, and remains valid also for the Minkowski
structure metric $q_6'$.  

Thus, Minkowski tensors and structure metrics both provide a ``geometrization''
of the bond-orientational order for spherical particles.  This suggests a strategy
to generalize bond-orientational order parameters towards aspherical particles,
such as ellipsoids, using generalized Voronoi tessellations and the $q_l'$.
Even applications to non-cellular shapes with arbitrary topology are possible,
albeit with altered interpretation \cite{Langmuir11_SchroederTurk,MickelSchroederTurkMecke:2012}.

Finally, our analysis supports the more frequent use of the low-weight $q_l$,
in particular $q_2'$, that have been largely overlooked in the literature. $q_2'$
carries the same information as the anisotropy index $\beta_1^{0,2}$ of
Refs.~\cite{EPL10_SchroederTurk,
KapferMickelSchroederTurkMecke:2012a,JSTAT10_Kapfer} (cf.~Fig.~\ref{fig:StructureDiagram}).
Both $q_2'$ and $\beta_1^{0,2}$ can be used to robustly classify collective states in particulate
matter according to their structural features.  Furthermore, $q_2'$ is very strongly
discerns between disordered configurations and such of high symmetry,
such as hcp, fcc, bcc, simple cubic, and icosahedral order.

Clearly, 30 years after the seminal publication by Steinhardt {\em et al.}, the
need for quantitative local structure analysis is more evident than ever. The
present paper reaffirms the validity and usefulness of the multipole expansion
method.  We have, however, described an amended version of the bond-orientational
order parameters that not only renders this method robust and uniquely defined,
but also gives a firmer interpretation of their geometric meaning.

\subsection*{Acknowledgments}
We are grateful to Tomaso Aste for the jammed LS data sets, to Shigenori Matsumoto, Tomoaki Nogawa,
Takashi Shimada, and Nobuyasu Ito for the MD data, to Markus Spanner for MC data, and to the authors of
Ref.~\cite{SkogeDonevStillingerTorquato:2006} for publishing their
Lubachevsky-Stillinger implementation. We thank Jean-Louis Barrat for his
suggestion to perform this study of $q_6$, and to Frank Rietz for comments
on the manuscript. We acknowledge support by the DFG through the
research group ``Geometry \& Physics of Spatial Random Systems'' under grants
SCHR 1148/3-1 and ME~1361/12-1.


\begin{thebibliography}{76}%
\makeatletter
\providecommand \@ifxundefined [1]{%
 \@ifx{#1\undefined}
}%
\providecommand \@ifnum [1]{%
 \ifnum #1\expandafter \@firstoftwo
 \else \expandafter \@secondoftwo
 \fi
}%
\providecommand \@ifx [1]{%
 \ifx #1\expandafter \@firstoftwo
 \else \expandafter \@secondoftwo
 \fi
}%
\providecommand \natexlab [1]{#1}%
\providecommand \enquote  [1]{``#1''}%
\providecommand \bibnamefont  [1]{#1}%
\providecommand \bibfnamefont [1]{#1}%
\providecommand \citenamefont [1]{#1}%
\providecommand \href@noop [0]{\@secondoftwo}%
\providecommand \href [0]{\begingroup \@sanitize@url \@href}%
\providecommand \@href[1]{\@@startlink{#1}\@@href}%
\providecommand \@@href[1]{\endgroup#1\@@endlink}%
\providecommand \@sanitize@url [0]{\catcode `\\12\catcode `\$12\catcode
  `\&12\catcode `\#12\catcode `\^12\catcode `\_12\catcode `\%12\relax}%
\providecommand \@@startlink[1]{}%
\providecommand \@@endlink[0]{}%
\providecommand \url  [0]{\begingroup\@sanitize@url \@url }%
\providecommand \@url [1]{\endgroup\@href {#1}{\urlprefix }}%
\providecommand \urlprefix  [0]{URL }%
\providecommand \Eprint [0]{\href }%
\providecommand \doibase [0]{http://dx.doi.org/}%
\providecommand \selectlanguage [0]{\@gobble}%
\providecommand \bibinfo  [0]{\@secondoftwo}%
\providecommand \bibfield  [0]{\@secondoftwo}%
\providecommand \translation [1]{[#1]}%
\providecommand \BibitemOpen [0]{}%
\providecommand \bibitemStop [0]{}%
\providecommand \bibitemNoStop [0]{.\EOS\space}%
\providecommand \EOS [0]{\spacefactor3000\relax}%
\providecommand \BibitemShut  [1]{\csname bibitem#1\endcsname}%
\let\auto@bib@innerbib\@empty
\bibitem [{\citenamefont {Steinhardt}, \citenamefont {Nelson},\ and\
  \citenamefont {Ronchetti}(1983)}]{PRB83_Steinhardt}%
  \BibitemOpen
  \bibfield  {author} {\bibinfo {author} {\bibfnamefont {P.}~\bibnamefont
  {Steinhardt}}, \bibinfo {author} {\bibfnamefont {D.}~\bibnamefont {Nelson}},
  \ and\ \bibinfo {author} {\bibfnamefont {M.}~\bibnamefont {Ronchetti}},\
  }\href@noop {} {\bibfield  {journal} {\bibinfo  {journal} {Phys.~Rev.~B.}\
  }\textbf {\bibinfo {volume} {28}},\ \bibinfo {pages} {784} (\bibinfo {year}
  {1983})}\BibitemShut {NoStop}%
\bibitem [{\citenamefont {Nelson}\ and\ \citenamefont
  {Halperin}(1979)}]{PRB79_Nelson}%
  \BibitemOpen
  \bibfield  {author} {\bibinfo {author} {\bibfnamefont {D.}~\bibnamefont
  {Nelson}}\ and\ \bibinfo {author} {\bibfnamefont {B.}~\bibnamefont
  {Halperin}},\ }\href@noop {} {\bibfield  {journal} {\bibinfo  {journal}
  {Phys.~Rev.~B}\ }\textbf {\bibinfo {volume} {19}},\ \bibinfo {pages} {2457}
  (\bibinfo {year} {1979})}\BibitemShut {NoStop}%
\bibitem [{\citenamefont {ten Wolde}, \citenamefont {Ruiz-Montero},\ and\
  \citenamefont {Frenkel}(1995)}]{PRL95_tenWolde}%
  \BibitemOpen
  \bibfield  {author} {\bibinfo {author} {\bibfnamefont {P.}~\bibnamefont {ten
  Wolde}}, \bibinfo {author} {\bibfnamefont {M.}~\bibnamefont {Ruiz-Montero}},
  \ and\ \bibinfo {author} {\bibfnamefont {D.}~\bibnamefont {Frenkel}},\
  }\href@noop {} {\bibfield  {journal} {\bibinfo  {journal} {Phys.~Rev.~Lett.}\
  }\textbf {\bibinfo {volume} {75}},\ \bibinfo {pages} {2714} (\bibinfo {year}
  {1995})}\BibitemShut {NoStop}%
\bibitem [{\citenamefont {Ni}\ and\ \citenamefont
  {Dijkstra}(2011)}]{JChemPhys11_Ni}%
  \BibitemOpen
  \bibfield  {author} {\bibinfo {author} {\bibfnamefont {R.}~\bibnamefont
  {Ni}}\ and\ \bibinfo {author} {\bibfnamefont {M.}~\bibnamefont {Dijkstra}},\
  }\href@noop {} {\bibfield  {journal} {\bibinfo  {journal} {J.~Chem.~Phys.}\
  }\textbf {\bibinfo {volume} {134}},\ \bibinfo {pages} {034501} (\bibinfo
  {year} {2011})}\BibitemShut {NoStop}%
\bibitem [{\citenamefont {Xu}, \citenamefont {Sun},\ and\ \citenamefont
  {An}(2010)}]{EPJE10_Xu}%
  \BibitemOpen
  \bibfield  {author} {\bibinfo {author} {\bibfnamefont {W.-S.}\ \bibnamefont
  {Xu}}, \bibinfo {author} {\bibfnamefont {Z.-Y.}\ \bibnamefont {Sun}}, \ and\
  \bibinfo {author} {\bibfnamefont {L.-J.}\ \bibnamefont {An}},\ }\href@noop {}
  {\bibfield  {journal} {\bibinfo  {journal} {Eur.~Phys.~J.~E}\ }\textbf
  {\bibinfo {volume} {31}},\ \bibinfo {pages} {377} (\bibinfo {year}
  {2010})}\BibitemShut {NoStop}%
\bibitem [{\citenamefont {Lechner}\ and\ \citenamefont
  {Dellago}(2008)}]{JChemPhys08_Lechner}%
  \BibitemOpen
  \bibfield  {author} {\bibinfo {author} {\bibfnamefont {W.}~\bibnamefont
  {Lechner}}\ and\ \bibinfo {author} {\bibfnamefont {C.}~\bibnamefont
  {Dellago}},\ }\href@noop {} {\bibfield  {journal} {\bibinfo  {journal}
  {J.~Chem.~Phys.}\ }\textbf {\bibinfo {volume} {129}},\ \bibinfo {pages}
  {114707} (\bibinfo {year} {2008})}\BibitemShut {NoStop}%
\bibitem [{\citenamefont {Valdes}\ \emph {et~al.}(2009)\citenamefont {Valdes},
  \citenamefont {Affouard}, \citenamefont {Descamps},\ and\ \citenamefont
  {Habasaki}}]{JChemPhys09_Valdes}%
  \BibitemOpen
  \bibfield  {author} {\bibinfo {author} {\bibfnamefont {L.-C.}\ \bibnamefont
  {Valdes}}, \bibinfo {author} {\bibfnamefont {F.}~\bibnamefont {Affouard}},
  \bibinfo {author} {\bibfnamefont {M.}~\bibnamefont {Descamps}}, \ and\
  \bibinfo {author} {\bibfnamefont {J.}~\bibnamefont {Habasaki}},\ }\href@noop
  {} {\bibfield  {journal} {\bibinfo  {journal} {J.~Chem.~Phys.}\ }\textbf
  {\bibinfo {volume} {130}},\ \bibinfo {pages} {154505} (\bibinfo {year}
  {2009})}\BibitemShut {NoStop}%
\bibitem [{\citenamefont {Kawasaki}\ and\ \citenamefont
  {Tanaka}(2010)}]{JPCM10_Kawasaki}%
  \BibitemOpen
  \bibfield  {author} {\bibinfo {author} {\bibfnamefont {T.}~\bibnamefont
  {Kawasaki}}\ and\ \bibinfo {author} {\bibfnamefont {H.}~\bibnamefont
  {Tanaka}},\ }\href@noop {} {\bibfield  {journal} {\bibinfo  {journal}
  {J.~Phys.:~Condens.~Matter}\ }\textbf {\bibinfo {volume} {22}},\ \bibinfo
  {pages} {232102} (\bibinfo {year} {2010})}\BibitemShut {NoStop}%
\bibitem [{\citenamefont {Wang}\ and\ \citenamefont
  {Gould}(2007)}]{PRE07_Wang}%
  \BibitemOpen
  \bibfield  {author} {\bibinfo {author} {\bibfnamefont {H.}~\bibnamefont
  {Wang}}\ and\ \bibinfo {author} {\bibfnamefont {H.}~\bibnamefont {Gould}},\
  }\href@noop {} {\bibfield  {journal} {\bibinfo  {journal} {Phys.~Rev.~E}\
  }\textbf {\bibinfo {volume} {76}},\ \bibinfo {pages} {031604} (\bibinfo
  {year} {2007})}\BibitemShut {NoStop}%
\bibitem [{\citenamefont {Wang}, \citenamefont {Teitel},\ and\ \citenamefont
  {Dellago}(2005)}]{JChemPhys05_Wang}%
  \BibitemOpen
  \bibfield  {author} {\bibinfo {author} {\bibfnamefont {Y.}~\bibnamefont
  {Wang}}, \bibinfo {author} {\bibfnamefont {S.}~\bibnamefont {Teitel}}, \ and\
  \bibinfo {author} {\bibfnamefont {C.}~\bibnamefont {Dellago}},\ }\href@noop
  {} {\bibfield  {journal} {\bibinfo  {journal} {J.~Chem.~Phys.}\ }\textbf
  {\bibinfo {volume} {122}},\ \bibinfo {pages} {214722} (\bibinfo {year}
  {2005})}\BibitemShut {NoStop}%
\bibitem [{\citenamefont {Keys}\ and\ \citenamefont
  {Glotzer}(2007)}]{PRL07_Keys}%
  \BibitemOpen
  \bibfield  {author} {\bibinfo {author} {\bibfnamefont {A.}~\bibnamefont
  {Keys}}\ and\ \bibinfo {author} {\bibfnamefont {S.}~\bibnamefont {Glotzer}},\
  }\href@noop {} {\bibfield  {journal} {\bibinfo  {journal} {Phys.~Rev.~Lett.}\
  }\textbf {\bibinfo {volume} {99}},\ \bibinfo {pages} {235503} (\bibinfo
  {year} {2007})}\BibitemShut {NoStop}%
\bibitem [{\citenamefont {Iacovella}\ \emph {et~al.}(2007)\citenamefont
  {Iacovella}, \citenamefont {Keys}, \citenamefont {Horsch},\ and\
  \citenamefont {Glotzer}}]{PRE07_Iacovella}%
  \BibitemOpen
  \bibfield  {author} {\bibinfo {author} {\bibfnamefont {C.}~\bibnamefont
  {Iacovella}}, \bibinfo {author} {\bibfnamefont {A.}~\bibnamefont {Keys}},
  \bibinfo {author} {\bibfnamefont {M.}~\bibnamefont {Horsch}}, \ and\ \bibinfo
  {author} {\bibfnamefont {S.}~\bibnamefont {Glotzer}},\ }\href@noop {}
  {\bibfield  {journal} {\bibinfo  {journal} {Phys.~Rev.~E}\ }\textbf {\bibinfo
  {volume} {75}},\ \bibinfo {pages} {040801} (\bibinfo {year}
  {2007})}\BibitemShut {NoStop}%
\bibitem [{\citenamefont {Chakravarty}, \citenamefont {Debenedetti},\ and\
  \citenamefont {Stillinger}(2007)}]{JChemPhys07_Chakravarty}%
  \BibitemOpen
  \bibfield  {author} {\bibinfo {author} {\bibfnamefont {C.}~\bibnamefont
  {Chakravarty}}, \bibinfo {author} {\bibfnamefont {P.~G.}\ \bibnamefont
  {Debenedetti}}, \ and\ \bibinfo {author} {\bibfnamefont {F.~H.}\ \bibnamefont
  {Stillinger}},\ }\href@noop {} {\bibfield  {journal} {\bibinfo  {journal}
  {J.~Chem.~Phys.}\ }\textbf {\bibinfo {volume} {126}},\ \bibinfo {pages}
  {204508} (\bibinfo {year} {2007})}\BibitemShut {NoStop}%
\bibitem [{\citenamefont {Calvo}\ and\ \citenamefont
  {Wales}(2009)}]{JChemPhys09_Calvo}%
  \BibitemOpen
  \bibfield  {author} {\bibinfo {author} {\bibfnamefont {F.}~\bibnamefont
  {Calvo}}\ and\ \bibinfo {author} {\bibfnamefont {D.~J.}\ \bibnamefont
  {Wales}},\ }\href@noop {} {\bibfield  {journal} {\bibinfo  {journal}
  {J.~Chem.~Phys.}\ }\textbf {\bibinfo {volume} {131}},\ \bibinfo {pages}
  {134504} (\bibinfo {year} {2009})}\BibitemShut {NoStop}%
\bibitem [{\citenamefont {Hern{\'{a}}ndez-Guzm{\'{a}}n}\ and\ \citenamefont
  {Weeks}(2009)}]{PNAS09-Hernandez-Guzman}%
  \BibitemOpen
  \bibfield  {author} {\bibinfo {author} {\bibfnamefont {J.}~\bibnamefont
  {Hern{\'{a}}ndez-Guzm{\'{a}}n}}\ and\ \bibinfo {author} {\bibfnamefont
  {E.~R.}\ \bibnamefont {Weeks}},\ }\href@noop {} {\bibfield  {journal}
  {\bibinfo  {journal} {Proc. Natl. Acad. Sci. U.S.A.}\ }\textbf {\bibinfo
  {volume} {106}},\ \bibinfo {pages} {15198} (\bibinfo {year}
  {2009})}\BibitemShut {NoStop}%
\bibitem [{\citenamefont {Binder}\ and\ \citenamefont {Kob}(2011)}]{KobBinder}%
  \BibitemOpen
  \bibfield  {author} {\bibinfo {author} {\bibfnamefont {K.}~\bibnamefont
  {Binder}}\ and\ \bibinfo {author} {\bibfnamefont {W.}~\bibnamefont {Kob}},\
  }\href@noop {} {\emph {\bibinfo {title} {Glassy Materials and Disordered
  Solids: An Introduction to Their Statistical Mechanics (Revised Edition)}}}\
  (\bibinfo  {publisher} {World Scientific Pub.~Co.},\ \bibinfo {year}
  {2011})\BibitemShut {NoStop}%
\bibitem [{\citenamefont {Ikeda}\ and\ \citenamefont
  {Miyazaki}(2011)}]{PRL11_Ikeda}%
  \BibitemOpen
  \bibfield  {author} {\bibinfo {author} {\bibfnamefont {A.}~\bibnamefont
  {Ikeda}}\ and\ \bibinfo {author} {\bibfnamefont {K.}~\bibnamefont
  {Miyazaki}},\ }\href@noop {} {\bibfield  {journal} {\bibinfo  {journal}
  {Phys.~Rev.~Lett.}\ }\textbf {\bibinfo {volume} {106}},\ \bibinfo {pages}
  {015701} (\bibinfo {year} {2011})}\BibitemShut {NoStop}%
\bibitem [{\citenamefont {Mokshin}\ and\ \citenamefont
  {Barrat}(2009)}]{JChemPhys09_Mokshin}%
  \BibitemOpen
  \bibfield  {author} {\bibinfo {author} {\bibfnamefont {A.~V.}\ \bibnamefont
  {Mokshin}}\ and\ \bibinfo {author} {\bibfnamefont {J.-L.}\ \bibnamefont
  {Barrat}},\ }\href@noop {} {\bibfield  {journal} {\bibinfo  {journal}
  {J.~Chem.~Phys.}\ }\textbf {\bibinfo {volume} {130}},\ \bibinfo {pages}
  {034502} (\bibinfo {year} {2009})}\BibitemShut {NoStop}%
\bibitem [{\citenamefont {Tanaka}\ \emph {et~al.}(2010)\citenamefont {Tanaka},
  \citenamefont {Kawasaki}, \citenamefont {Shintani},\ and\ \citenamefont
  {Watanabe}}]{NatMat10_Tanaka}%
  \BibitemOpen
  \bibfield  {author} {\bibinfo {author} {\bibfnamefont {H.}~\bibnamefont
  {Tanaka}}, \bibinfo {author} {\bibfnamefont {T.}~\bibnamefont {Kawasaki}},
  \bibinfo {author} {\bibfnamefont {H.}~\bibnamefont {Shintani}}, \ and\
  \bibinfo {author} {\bibfnamefont {K.}~\bibnamefont {Watanabe}},\ }\href@noop
  {} {\bibfield  {journal} {\bibinfo  {journal} {Nature Mater.}\ }\textbf
  {\bibinfo {volume} {9}},\ \bibinfo {pages} {324} (\bibinfo {year}
  {2010})}\BibitemShut {NoStop}%
\bibitem [{\citenamefont {Lochmann}\ \emph {et~al.}(2006)\citenamefont
  {Lochmann}, \citenamefont {Anikeenko}, \citenamefont {Elsner}, \citenamefont
  {Medvedev},\ and\ \citenamefont {Stoyan}}]{EPJB06_Lochmann}%
  \BibitemOpen
  \bibfield  {author} {\bibinfo {author} {\bibfnamefont {K.}~\bibnamefont
  {Lochmann}}, \bibinfo {author} {\bibfnamefont {A.}~\bibnamefont {Anikeenko}},
  \bibinfo {author} {\bibfnamefont {A.}~\bibnamefont {Elsner}}, \bibinfo
  {author} {\bibfnamefont {N.}~\bibnamefont {Medvedev}}, \ and\ \bibinfo
  {author} {\bibfnamefont {D.}~\bibnamefont {Stoyan}},\ }\href@noop {}
  {\bibfield  {journal} {\bibinfo  {journal} {Eur.~Phys.~J.~B}\ }\textbf
  {\bibinfo {volume} {53}},\ \bibinfo {pages} {67} (\bibinfo {year}
  {2006})}\BibitemShut {NoStop}%
\bibitem [{\citenamefont {Schilling}\ \emph {et~al.}(2010)\citenamefont
  {Schilling}, \citenamefont {Sch{\"{o}}pe}, \citenamefont {Oettel},
  \citenamefont {Opletal},\ and\ \citenamefont {Snook}}]{PRL10_Schilling}%
  \BibitemOpen
  \bibfield  {author} {\bibinfo {author} {\bibfnamefont {T.}~\bibnamefont
  {Schilling}}, \bibinfo {author} {\bibfnamefont {H.}~\bibnamefont
  {Sch{\"{o}}pe}}, \bibinfo {author} {\bibfnamefont {M.}~\bibnamefont
  {Oettel}}, \bibinfo {author} {\bibfnamefont {G.}~\bibnamefont {Opletal}}, \
  and\ \bibinfo {author} {\bibfnamefont {I.}~\bibnamefont {Snook}},\
  }\href@noop {} {\bibfield  {journal} {\bibinfo  {journal} {Phys.~Rev.~Lett.}\
  }\textbf {\bibinfo {volume} {105}},\ \bibinfo {pages} {025701} (\bibinfo
  {year} {2010})}\BibitemShut {NoStop}%
\bibitem [{\citenamefont {van Duijneveldt}\ and\ \citenamefont
  {Frenkel}(1992)}]{JChemPhys91_vanDuijneveldt}%
  \BibitemOpen
  \bibfield  {author} {\bibinfo {author} {\bibfnamefont {J.~S.}\ \bibnamefont
  {van Duijneveldt}}\ and\ \bibinfo {author} {\bibfnamefont {D.}~\bibnamefont
  {Frenkel}},\ }\href@noop {} {\bibfield  {journal} {\bibinfo  {journal}
  {J.~Chem.~Phys.}\ }\textbf {\bibinfo {volume} {96}},\ \bibinfo {pages} {4655}
  (\bibinfo {year} {1992})}\BibitemShut {NoStop}%
\bibitem [{\citenamefont {Wouterse}\ and\ \citenamefont
  {Philipse}(2006)}]{JChemPhys06_Wouterse}%
  \BibitemOpen
  \bibfield  {author} {\bibinfo {author} {\bibfnamefont {A.}~\bibnamefont
  {Wouterse}}\ and\ \bibinfo {author} {\bibfnamefont {A.~P.}\ \bibnamefont
  {Philipse}},\ }\href@noop {} {\bibfield  {journal} {\bibinfo  {journal}
  {J.~Chem.~Phys.}\ }\textbf {\bibinfo {volume} {125}},\ \bibinfo {pages}
  {194709} (\bibinfo {year} {2006})}\BibitemShut {NoStop}%
\bibitem [{\citenamefont {Duff}\ and\ \citenamefont
  {Lacks}(2007)}]{PRE07_Duff}%
  \BibitemOpen
  \bibfield  {author} {\bibinfo {author} {\bibfnamefont {N.}~\bibnamefont
  {Duff}}\ and\ \bibinfo {author} {\bibfnamefont {D.}~\bibnamefont {Lacks}},\
  }\href@noop {} {\bibfield  {journal} {\bibinfo  {journal} {Phys.~Rev.~E}\
  }\textbf {\bibinfo {volume} {75}},\ \bibinfo {pages} {031501} (\bibinfo
  {year} {2007})}\BibitemShut {NoStop}%
\bibitem [{\citenamefont {Abraham}\ and\ \citenamefont
  {Bagchi}(2008)}]{PRE08_Abraham}%
  \BibitemOpen
  \bibfield  {author} {\bibinfo {author} {\bibfnamefont {S.}~\bibnamefont
  {Abraham}}\ and\ \bibinfo {author} {\bibfnamefont {B.}~\bibnamefont
  {Bagchi}},\ }\href@noop {} {\bibfield  {journal} {\bibinfo  {journal}
  {Phys.~Rev.~E}\ }\textbf {\bibinfo {volume} {78}},\ \bibinfo {pages} {051501}
  (\bibinfo {year} {2008})}\BibitemShut {NoStop}%
\bibitem [{Note1()}]{Note1}%
  \BibitemOpen
  \bibinfo {note} {This expression assumes that neighborhood is a symmetric
  concept, such that $a\in \protect \mathrm {NN}(b)$ implies that $b\in
  \protect \mathrm {NN}(a)$. This is correct for the definitions of
  neighborhood based on cutoff radii and on the Delaney triangulation, but not
  for the definition based on a fixed number of neighbors.}\BibitemShut {Stop}%
\bibitem [{\citenamefont {Kelchner}, \citenamefont {Plimpton},\ and\
  \citenamefont {Hamilton}(1998)}]{KelchnerPlimptonHamilton:1998}%
  \BibitemOpen
  \bibfield  {author} {\bibinfo {author} {\bibfnamefont {C.}~\bibnamefont
  {Kelchner}}, \bibinfo {author} {\bibfnamefont {S.}~\bibnamefont {Plimpton}},
  \ and\ \bibinfo {author} {\bibfnamefont {J.}~\bibnamefont {Hamilton}},\
  }\href@noop {} {\bibfield  {journal} {\bibinfo  {journal} {Phys.~Rev.~B}\
  }\textbf {\bibinfo {volume} {58}},\ \bibinfo {pages} {11085} (\bibinfo {year}
  {1998})}\BibitemShut {NoStop}%
\bibitem [{\citenamefont {Edwards}\ and\ \citenamefont
  {Grinev}(2001)}]{PhysicaA01_Edwards}%
  \BibitemOpen
  \bibfield  {author} {\bibinfo {author} {\bibfnamefont {S.}~\bibnamefont
  {Edwards}}\ and\ \bibinfo {author} {\bibfnamefont {D.}~\bibnamefont
  {Grinev}},\ }\href@noop {} {\bibfield  {journal} {\bibinfo  {journal}
  {Physica A}\ }\textbf {\bibinfo {volume} {302}},\ \bibinfo {pages} {162}
  (\bibinfo {year} {2001})}\BibitemShut {NoStop}%
\bibitem [{\citenamefont {Bargiel}\ and\ \citenamefont
  {Tory}(2001)}]{Bargiel:2001:0921-8831:533}%
  \BibitemOpen
  \bibfield  {author} {\bibinfo {author} {\bibfnamefont {M.}~\bibnamefont
  {Bargiel}}\ and\ \bibinfo {author} {\bibfnamefont {E.~M.}\ \bibnamefont
  {Tory}},\ }\href@noop {} {\bibfield  {journal} {\bibinfo  {journal}
  {Adv.~Powder Technol.}\ }\textbf {\bibinfo {volume} {12}},\ \bibinfo {pages}
  {533} (\bibinfo {year} {2001})}\BibitemShut {NoStop}%
\bibitem [{\citenamefont {Armstrong}\ \emph {et~al.}(2009)\citenamefont
  {Armstrong}, \citenamefont {Knieke}, \citenamefont {Mackovic}, \citenamefont
  {Frank}, \citenamefont {Hartmaier}, \citenamefont {Göken},\ and\
  \citenamefont {Peukert}}]{Armstrong20093060}%
  \BibitemOpen
  \bibfield  {author} {\bibinfo {author} {\bibfnamefont {P.}~\bibnamefont
  {Armstrong}}, \bibinfo {author} {\bibfnamefont {C.}~\bibnamefont {Knieke}},
  \bibinfo {author} {\bibfnamefont {M.}~\bibnamefont {Mackovic}}, \bibinfo
  {author} {\bibfnamefont {G.}~\bibnamefont {Frank}}, \bibinfo {author}
  {\bibfnamefont {A.}~\bibnamefont {Hartmaier}}, \bibinfo {author}
  {\bibfnamefont {M.}~\bibnamefont {Göken}}, \ and\ \bibinfo {author}
  {\bibfnamefont {W.}~\bibnamefont {Peukert}},\ }\href@noop {} {\bibfield
  {journal} {\bibinfo  {journal} {Acta Mat.}\ }\textbf {\bibinfo {volume}
  {57}},\ \bibinfo {pages} {3060 } (\bibinfo {year} {2009})}\BibitemShut
  {NoStop}%
\bibitem [{\citenamefont {Gray}\ and\ \citenamefont
  {Gubbins}(1984)}]{GrayGubbins}%
  \BibitemOpen
  \bibfield  {author} {\bibinfo {author} {\bibfnamefont {C.}~\bibnamefont
  {Gray}}\ and\ \bibinfo {author} {\bibfnamefont {K.}~\bibnamefont {Gubbins}},\
  }\href@noop {} {\emph {\bibinfo {title} {Theory of molecular fluids (Volume
  1: Fundamentals)}}}\ (\bibinfo  {publisher} {Clarendon Press, Oxford},\
  \bibinfo {year} {1984})\BibitemShut {NoStop}%
\bibitem [{\citenamefont {Wigner}(1959)}]{wigner1959gruppentheorie}%
  \BibitemOpen
  \bibfield  {author} {\bibinfo {author} {\bibfnamefont {E.}~\bibnamefont
  {Wigner}},\ }\href@noop {} {\emph {\bibinfo {title} {Gruppentheorie und ihre
  Anwendung auf die Quantenmechanik der Atomspektren}}},\ Pure and applied
  physics\ (\bibinfo  {publisher} {Academic Press},\ \bibinfo {year}
  {1959})\BibitemShut {NoStop}%
\bibitem [{Note2()}]{Note2}%
  \BibitemOpen
  \bibinfo {note} {Although we will not use the global bond order parameter
  $Q_l$ we define it for completeness as $ Q_{l}= \protect \sqrt {\protect
  \frac {4\pi }{2l+1}\DOTSB \sum@ \slimits@ _{m=-l}^{l}\left \delimiter
  "026A30C \protect \frac {1}{\protect \mathcal {N}} \DOTSB \sum@ \slimits@
  _{k=1}^N \DOTSB \sum@ \slimits@ _{j=1}^{n_a} Y_{lm}\left (\theta _j,\varphi
  _j\right ) \right \delimiter "026A30C ^2}, $ where $N$ is the number of
  spherical particles and $\protect \mathcal {N}=\DOTSB \sum@ \slimits@
  _{a=1}^N n(a)$ the number of all bonds. This is, the average over all bonds
  is taken inside the norm. For disordered systems the sum over the $Y_{lm}$
  vanishes as $\protect \mathcal {N}^{-1/2}$, while it remains finite for
  common crystalline structures \cite
  {PRB83_Steinhardt,JChemPhys96_Rintoul}.}\BibitemShut {Stop}%
\bibitem [{\citenamefont {Aste}, \citenamefont {Saadatfar},\ and\ \citenamefont
  {Senden}(2005)}]{PRE05_Aste}%
  \BibitemOpen
  \bibfield  {author} {\bibinfo {author} {\bibfnamefont {T.}~\bibnamefont
  {Aste}}, \bibinfo {author} {\bibfnamefont {M.}~\bibnamefont {Saadatfar}}, \
  and\ \bibinfo {author} {\bibfnamefont {T.}~\bibnamefont {Senden}},\
  }\href@noop {} {\bibfield  {journal} {\bibinfo  {journal} {Phys.~Rev.~E}\
  }\textbf {\bibinfo {volume} {71}},\ \bibinfo {pages} {061302} (\bibinfo
  {year} {2005})}\BibitemShut {NoStop}%
\bibitem [{\citenamefont {Klumov}(2011)}]{Uspekhi11_Klumov}%
  \BibitemOpen
  \bibfield  {author} {\bibinfo {author} {\bibfnamefont {B.~A.}\ \bibnamefont
  {Klumov}},\ }\href@noop {} {\bibfield  {journal} {\bibinfo  {journal}
  {Physics-Uspekhi}\ }\textbf {\bibinfo {volume} {53}},\ \bibinfo {pages}
  {1053} (\bibinfo {year} {2011})}\BibitemShut {NoStop}%
\bibitem [{\citenamefont {Yiannourakou}, \citenamefont {Economou},\ and\
  \citenamefont {Bitsanis}(2010)}]{JChemPhys10_Yiannourakou}%
  \BibitemOpen
  \bibfield  {author} {\bibinfo {author} {\bibfnamefont {M.}~\bibnamefont
  {Yiannourakou}}, \bibinfo {author} {\bibfnamefont {I.~G.}\ \bibnamefont
  {Economou}}, \ and\ \bibinfo {author} {\bibfnamefont {I.~A.}\ \bibnamefont
  {Bitsanis}},\ }\href@noop {} {\bibfield  {journal} {\bibinfo  {journal}
  {J.~Chem.~Phys.}\ }\textbf {\bibinfo {volume} {133}},\ \bibinfo {pages}
  {224901} (\bibinfo {year} {2010})}\BibitemShut {NoStop}%
\bibitem [{\citenamefont {Martin}(2008)}]{PRE08_Martin}%
  \BibitemOpen
  \bibfield  {author} {\bibinfo {author} {\bibfnamefont {C.~L.}\ \bibnamefont
  {Martin}},\ }\href@noop {} {\bibfield  {journal} {\bibinfo  {journal}
  {Phys.~Rev.~E}\ }\textbf {\bibinfo {volume} {77}},\ \bibinfo {pages} {031307}
  (\bibinfo {year} {2008})}\BibitemShut {NoStop}%
\bibitem [{\citenamefont {Kapfer}\ \emph
  {et~al.}(2012{\natexlab{a}})\citenamefont {Kapfer}, \citenamefont {Mickel},
  \citenamefont {Mecke},\ and\ \citenamefont
  {Schr\"oder-Turk}}]{KapferMickelSchroederTurkMecke:2012a}%
  \BibitemOpen
  \bibfield  {author} {\bibinfo {author} {\bibfnamefont {S.~C.}\ \bibnamefont
  {Kapfer}}, \bibinfo {author} {\bibfnamefont {W.}~\bibnamefont {Mickel}},
  \bibinfo {author} {\bibfnamefont {K.}~\bibnamefont {Mecke}}, \ and\ \bibinfo
  {author} {\bibfnamefont {G.~E.}\ \bibnamefont {Schr\"oder-Turk}},\
  }\href@noop {} {\bibfield  {journal} {\bibinfo  {journal} {Phys.~Rev.~E}\
  }\textbf {\bibinfo {volume} {85}},\ \bibinfo {pages} {030301} (\bibinfo
  {year} {2012}{\natexlab{a}})}\BibitemShut {NoStop}%
\bibitem [{Note3()}]{Note3}%
  \BibitemOpen
  \bibinfo {note} {Normalized bond order functions are $q_{lm}(a):=\left
  (\DOTSB \sum@ \slimits@ _{i=1}^{n(a)}Y_{lm}\right )/q_l(a)$ for particle $a$
  and the dot-product is $d_{ab}:=\DOTSB \sum@ \slimits@ _{m=-l}^{l}
  q_{lm}(a)q_{lm}^*(b)$ of spheres $a$ and $b$. A particle is defined as member
  of a solid-like cluster, if the dot-product with $n_0$ NN exceeds a certain
  threshold $d_0$.}\BibitemShut {Stop}%
\bibitem [{\citenamefont {Mokshin}\ and\ \citenamefont
  {Barrat}(2010)}]{PRE10_Mokshin}%
  \BibitemOpen
  \bibfield  {author} {\bibinfo {author} {\bibfnamefont {A.}~\bibnamefont
  {Mokshin}}\ and\ \bibinfo {author} {\bibfnamefont {J.-L.}\ \bibnamefont
  {Barrat}},\ }\href@noop {} {\bibfield  {journal} {\bibinfo  {journal}
  {Phys.~Rev.~E}\ }\textbf {\bibinfo {volume} {82}},\ \bibinfo {pages} {021505}
  (\bibinfo {year} {2010})}\BibitemShut {NoStop}%
\bibitem [{\citenamefont {Panaitescu}\ and\ \citenamefont
  {Kudrolli}(2010)}]{PRE10_Panaitescu}%
  \BibitemOpen
  \bibfield  {author} {\bibinfo {author} {\bibfnamefont {A.}~\bibnamefont
  {Panaitescu}}\ and\ \bibinfo {author} {\bibfnamefont {A.}~\bibnamefont
  {Kudrolli}},\ }\href@noop {} {\bibfield  {journal} {\bibinfo  {journal}
  {Phys.~Rev.~E}\ }\textbf {\bibinfo {volume} {81}},\ \bibinfo {pages}
  {060301(R)} (\bibinfo {year} {2010})}\BibitemShut {NoStop}%
\bibitem [{\citenamefont {de~Oliveira}\ \emph {et~al.}(2006)\citenamefont
  {de~Oliveira}, \citenamefont {Netz}, \citenamefont {Colla},\ and\
  \citenamefont {Barbosa}}]{JChemPhys06_deOliveira}%
  \BibitemOpen
  \bibfield  {author} {\bibinfo {author} {\bibfnamefont {A.~B.}\ \bibnamefont
  {de~Oliveira}}, \bibinfo {author} {\bibfnamefont {P.~A.}\ \bibnamefont
  {Netz}}, \bibinfo {author} {\bibfnamefont {T.}~\bibnamefont {Colla}}, \ and\
  \bibinfo {author} {\bibfnamefont {M.~C.}\ \bibnamefont {Barbosa}},\
  }\href@noop {} {\bibfield  {journal} {\bibinfo  {journal} {J.~Chem.~Phys.}\
  }\textbf {\bibinfo {volume} {125}},\ \bibinfo {pages} {124503} (\bibinfo
  {year} {2006})}\BibitemShut {NoStop}%
\bibitem [{\citenamefont {Yan}\ \emph {et~al.}(2007)\citenamefont {Yan},
  \citenamefont {Buldyrev}, \citenamefont {Kumar}, \citenamefont
  {Giovambattista}, \citenamefont {Debenedetti},\ and\ \citenamefont
  {Stanley}}]{PRE07_Yan}%
  \BibitemOpen
  \bibfield  {author} {\bibinfo {author} {\bibfnamefont {Z.}~\bibnamefont
  {Yan}}, \bibinfo {author} {\bibfnamefont {S.~V.}\ \bibnamefont {Buldyrev}},
  \bibinfo {author} {\bibfnamefont {P.}~\bibnamefont {Kumar}}, \bibinfo
  {author} {\bibfnamefont {N.}~\bibnamefont {Giovambattista}}, \bibinfo
  {author} {\bibfnamefont {P.}~\bibnamefont {Debenedetti}}, \ and\ \bibinfo
  {author} {\bibfnamefont {H.}~\bibnamefont {Stanley}},\ }\href@noop {}
  {\bibfield  {journal} {\bibinfo  {journal} {Phys.~Rev.~E}\ }\textbf {\bibinfo
  {volume} {76}},\ \bibinfo {pages} {051201} (\bibinfo {year}
  {2007})}\BibitemShut {NoStop}%
\bibitem [{\citenamefont {Wallace}\ and\ \citenamefont
  {Jo{\'{o}}s}(2006)}]{PRL06_Wallace}%
  \BibitemOpen
  \bibfield  {author} {\bibinfo {author} {\bibfnamefont {M.}~\bibnamefont
  {Wallace}}\ and\ \bibinfo {author} {\bibfnamefont {B.}~\bibnamefont
  {Jo{\'{o}}s}},\ }\href@noop {} {\bibfield  {journal} {\bibinfo  {journal}
  {Phys.~Rev.~Lett.}\ }\textbf {\bibinfo {volume} {96}},\ \bibinfo {pages}
  {025501} (\bibinfo {year} {2006})}\BibitemShut {NoStop}%
\bibitem [{\citenamefont {Kansal}, \citenamefont {Torquato},\ and\
  \citenamefont {Stillinger}(2002)}]{PRE02_Kansal}%
  \BibitemOpen
  \bibfield  {author} {\bibinfo {author} {\bibfnamefont {A.}~\bibnamefont
  {Kansal}}, \bibinfo {author} {\bibfnamefont {S.}~\bibnamefont {Torquato}}, \
  and\ \bibinfo {author} {\bibfnamefont {F.}~\bibnamefont {Stillinger}},\
  }\href@noop {} {\bibfield  {journal} {\bibinfo  {journal} {Phys.~Rev.~E}\
  }\textbf {\bibinfo {volume} {66}},\ \bibinfo {pages} {041109} (\bibinfo
  {year} {2002})}\BibitemShut {NoStop}%
\bibitem [{\citenamefont {Errington}\ and\ \citenamefont
  {Debenedetti}(2001)}]{Natur01_Errington}%
  \BibitemOpen
  \bibfield  {author} {\bibinfo {author} {\bibfnamefont {J.~R.}\ \bibnamefont
  {Errington}}\ and\ \bibinfo {author} {\bibfnamefont {P.~G.}\ \bibnamefont
  {Debenedetti}},\ }\href@noop {} {\bibfield  {journal} {\bibinfo  {journal}
  {Nature}\ }\textbf {\bibinfo {volume} {409}},\ \bibinfo {pages} {318}
  (\bibinfo {year} {2001})}\BibitemShut {NoStop}%
\bibitem [{\citenamefont {Odriozola}(2009)}]{JChemPhys09_Odriozola}%
  \BibitemOpen
  \bibfield  {author} {\bibinfo {author} {\bibfnamefont {G.}~\bibnamefont
  {Odriozola}},\ }\href@noop {} {\bibfield  {journal} {\bibinfo  {journal}
  {J.~Chem.~Phys.}\ }\textbf {\bibinfo {volume} {131}},\ \bibinfo {pages}
  {144107} (\bibinfo {year} {2009})}\BibitemShut {NoStop}%
\bibitem [{\citenamefont {Kurita}\ and\ \citenamefont
  {Weeks}(2010)}]{PRE10_Kurita}%
  \BibitemOpen
  \bibfield  {author} {\bibinfo {author} {\bibfnamefont {R.}~\bibnamefont
  {Kurita}}\ and\ \bibinfo {author} {\bibfnamefont {E.}~\bibnamefont {Weeks}},\
  }\href@noop {} {\bibfield  {journal} {\bibinfo  {journal} {Phys.~Rev.~E}\
  }\textbf {\bibinfo {volume} {82}},\ \bibinfo {pages} {011403} (\bibinfo
  {year} {2010})}\BibitemShut {NoStop}%
\bibitem [{\citenamefont {Barber}, \citenamefont {Dobkin},\ and\ \citenamefont
  {Huhdanpaa}(1996)}]{BarberDobkinHuhdanpaa:1996}%
  \BibitemOpen
  \bibfield  {author} {\bibinfo {author} {\bibfnamefont {C.~B.}\ \bibnamefont
  {Barber}}, \bibinfo {author} {\bibfnamefont {D.~P.}\ \bibnamefont {Dobkin}},
  \ and\ \bibinfo {author} {\bibfnamefont {H.}~\bibnamefont {Huhdanpaa}},\
  }\href@noop {} {\bibfield  {journal} {\bibinfo  {journal} {ACM
  Trans.~Math.~Softw.}\ }\textbf {\bibinfo {volume} {22}},\ \bibinfo {pages}
  {469} (\bibinfo {year} {1996})}\BibitemShut {NoStop}%
\bibitem [{Note4()}]{Note4}%
  \BibitemOpen
  \bibinfo {note} {The definition of NN via the Delaunay graph is equivalent to
  the definition via Voronoi neighbors: spheres share a Delaunay edge, whenever
  their respective Voronoi cells have a shared facet (regardless of the area of
  the Voronoi facet).}\BibitemShut {Stop}%
\bibitem [{\citenamefont {Senthil~Kumar}\ and\ \citenamefont
  {Kumaran}(2006)}]{JChemPhys06_Kumar}%
  \BibitemOpen
  \bibfield  {author} {\bibinfo {author} {\bibfnamefont {V.}~\bibnamefont
  {Senthil~Kumar}}\ and\ \bibinfo {author} {\bibfnamefont {V.}~\bibnamefont
  {Kumaran}},\ }\href@noop {} {\bibfield  {journal} {\bibinfo  {journal}
  {J.~Chem.~Phys.}\ }\textbf {\bibinfo {volume} {124}},\ \bibinfo {pages}
  {204508} (\bibinfo {year} {2006})}\BibitemShut {NoStop}%
\bibitem [{\citenamefont {{Matsumoto}}\ \emph {et~al.}(2010)\citenamefont
  {{Matsumoto}}, \citenamefont {{Nogawa}}, \citenamefont {{Shimada}},\ and\
  \citenamefont {{Ito}}}]{arxiv10:Nogawa}%
  \BibitemOpen
  \bibfield  {author} {\bibinfo {author} {\bibfnamefont {S.}~\bibnamefont
  {{Matsumoto}}}, \bibinfo {author} {\bibfnamefont {T.}~\bibnamefont
  {{Nogawa}}}, \bibinfo {author} {\bibfnamefont {T.}~\bibnamefont {{Shimada}}},
  \ and\ \bibinfo {author} {\bibfnamefont {N.}~\bibnamefont {{Ito}}},\
  }\href@noop {} {\bibfield  {journal} {\bibinfo  {journal} {ArXiv e-prints}\ }
  (\bibinfo {year} {2010})},\ \Eprint {http://arxiv.org/abs/1005.4295}
  {1005.4295} \BibitemShut {NoStop}%
\bibitem [{\citenamefont {Kapfer}\ \emph {et~al.}(2010)\citenamefont {Kapfer},
  \citenamefont {Mickel}, \citenamefont {Schaller}, \citenamefont {Spanner},
  \citenamefont {Goll}, \citenamefont {Nogawa}, \citenamefont {Ito},
  \citenamefont {Mecke},\ and\ \citenamefont
  {Schr{\"{o}}der-Turk}}]{JSTAT10_Kapfer}%
  \BibitemOpen
  \bibfield  {author} {\bibinfo {author} {\bibfnamefont {S.~C.}\ \bibnamefont
  {Kapfer}}, \bibinfo {author} {\bibfnamefont {W.}~\bibnamefont {Mickel}},
  \bibinfo {author} {\bibfnamefont {F.~M.}\ \bibnamefont {Schaller}}, \bibinfo
  {author} {\bibfnamefont {M.}~\bibnamefont {Spanner}}, \bibinfo {author}
  {\bibfnamefont {C.}~\bibnamefont {Goll}}, \bibinfo {author} {\bibfnamefont
  {T.}~\bibnamefont {Nogawa}}, \bibinfo {author} {\bibfnamefont
  {N.}~\bibnamefont {Ito}}, \bibinfo {author} {\bibfnamefont {K.}~\bibnamefont
  {Mecke}}, \ and\ \bibinfo {author} {\bibfnamefont {G.~E.}\ \bibnamefont
  {Schr{\"{o}}der-Turk}},\ }\href@noop {} {\bibfield  {journal} {\bibinfo
  {journal} {J.~Stat.~Mech.~Theor.~Exp.}\ }\textbf {\bibinfo {volume} {2010}},\
  \bibinfo {pages} {P11010} (\bibinfo {year} {2010})}\BibitemShut {NoStop}%
\bibitem [{Note5()}]{Note5}%
  \BibitemOpen
  \bibinfo {note} {Event driven MD simulations to explore the super-cooled
  regime use the Matsumoto algorithm from Ref.~\cite {arxiv10:Nogawa}. In this
  algorithm, spheres are expanded until they touch the closest Voronoi facet or
  until they reach the final radius. This creates a transient polydisperse
  ensemble, which is relaxed by thermal motion, followed by an expansion step.
  This procedure is iterated until a monodisperse HS system at predefined
  packing fraction is obtained.}\BibitemShut {Stop}%
\bibitem [{\citenamefont {Woodcock}(1997)}]{Nature97_Woodcock}%
  \BibitemOpen
  \bibfield  {author} {\bibinfo {author} {\bibfnamefont {L.~V.}\ \bibnamefont
  {Woodcock}},\ }\href@noop {} {\bibfield  {journal} {\bibinfo  {journal}
  {Nature}\ }\textbf {\bibinfo {volume} {385}},\ \bibinfo {pages} {141}
  (\bibinfo {year} {1997})}\BibitemShut {NoStop}%
\bibitem [{\citenamefont {Rein~ten Wolde}, \citenamefont {Ruiz-Montero},\ and\
  \citenamefont {Frenkel}(1996)}]{JChemPhys96_tenWolde}%
  \BibitemOpen
  \bibfield  {author} {\bibinfo {author} {\bibfnamefont {P.}~\bibnamefont
  {Rein~ten Wolde}}, \bibinfo {author} {\bibfnamefont {M.~J.}\ \bibnamefont
  {Ruiz-Montero}}, \ and\ \bibinfo {author} {\bibfnamefont {D.}~\bibnamefont
  {Frenkel}},\ }\href@noop {} {\bibfield  {journal} {\bibinfo  {journal}
  {J.~Chem.~Phys.}\ }\textbf {\bibinfo {volume} {104}},\ \bibinfo {pages}
  {9932} (\bibinfo {year} {1996})}\BibitemShut {NoStop}%
\bibitem [{\citenamefont {Troadec}, \citenamefont {Gervois},\ and\
  \citenamefont {Oger}(2007)}]{EPL98_Troadec}%
  \BibitemOpen
  \bibfield  {author} {\bibinfo {author} {\bibfnamefont {J.~P.}\ \bibnamefont
  {Troadec}}, \bibinfo {author} {\bibfnamefont {A.}~\bibnamefont {Gervois}}, \
  and\ \bibinfo {author} {\bibfnamefont {L.}~\bibnamefont {Oger}},\ }\href
  {\doibase 10.1209/epl/i1998-00224-x} {\bibfield  {journal} {\bibinfo
  {journal} {Europhysics Letters (EPL)}\ }\textbf {\bibinfo {volume} {42}},\
  \bibinfo {pages} {167} (\bibinfo {year} {2007})}\BibitemShut {NoStop}%
\bibitem [{\citenamefont {Rintoul}\ and\ \citenamefont
  {Torquato}(1996)}]{JChemPhys96_Rintoul}%
  \BibitemOpen
  \bibfield  {author} {\bibinfo {author} {\bibfnamefont {M.~D.}\ \bibnamefont
  {Rintoul}}\ and\ \bibinfo {author} {\bibfnamefont {S.}~\bibnamefont
  {Torquato}},\ }\href@noop {} {\bibfield  {journal} {\bibinfo  {journal}
  {J.~Chem.~Phys.}\ }\textbf {\bibinfo {volume} {105}},\ \bibinfo {pages}
  {9258} (\bibinfo {year} {1996})}\BibitemShut {NoStop}%
\bibitem [{\citenamefont {Schr{\"o}der-Turk}\ \emph
  {et~al.}(2011{\natexlab{a}})\citenamefont {Schr{\"o}der-Turk}, \citenamefont
  {Mickel}, \citenamefont {Kapfer}, \citenamefont {Klatt}, \citenamefont
  {Schaller}, \citenamefont {Hoffmann}, \citenamefont {Kleppmann},
  \citenamefont {Armstrong}, \citenamefont {Inayat}, \citenamefont {Hug},
  \citenamefont {Reichelsdorfer}, \citenamefont {Peukert}, \citenamefont
  {Schwieger},\ and\ \citenamefont {Mecke}}]{AdvMatt11_SchroederTurk}%
  \BibitemOpen
  \bibfield  {author} {\bibinfo {author} {\bibfnamefont {G.~E.}\ \bibnamefont
  {Schr{\"o}der-Turk}}, \bibinfo {author} {\bibfnamefont {W.}~\bibnamefont
  {Mickel}}, \bibinfo {author} {\bibfnamefont {S.~C.}\ \bibnamefont {Kapfer}},
  \bibinfo {author} {\bibfnamefont {M.~A.}\ \bibnamefont {Klatt}}, \bibinfo
  {author} {\bibfnamefont {F.~M.}\ \bibnamefont {Schaller}}, \bibinfo {author}
  {\bibfnamefont {M.~J.~F.}\ \bibnamefont {Hoffmann}}, \bibinfo {author}
  {\bibfnamefont {N.}~\bibnamefont {Kleppmann}}, \bibinfo {author}
  {\bibfnamefont {P.}~\bibnamefont {Armstrong}}, \bibinfo {author}
  {\bibfnamefont {A.}~\bibnamefont {Inayat}}, \bibinfo {author} {\bibfnamefont
  {D.}~\bibnamefont {Hug}}, \bibinfo {author} {\bibfnamefont {M.}~\bibnamefont
  {Reichelsdorfer}}, \bibinfo {author} {\bibfnamefont {W.}~\bibnamefont
  {Peukert}}, \bibinfo {author} {\bibfnamefont {W.}~\bibnamefont {Schwieger}},
  \ and\ \bibinfo {author} {\bibfnamefont {K.}~\bibnamefont {Mecke}},\
  }\href@noop {} {\bibfield  {journal} {\bibinfo  {journal} {Adv.~Mater.}\
  }\textbf {\bibinfo {volume} {23}},\ \bibinfo {pages} {2535} (\bibinfo {year}
  {2011}{\natexlab{a}})}\BibitemShut {NoStop}%
\bibitem [{\citenamefont {Schneider}\ and\ \citenamefont
  {Weil}(2000)}]{SchneiderWeil}%
  \BibitemOpen
  \bibfield  {author} {\bibinfo {author} {\bibfnamefont {R.}~\bibnamefont
  {Schneider}}\ and\ \bibinfo {author} {\bibfnamefont {W.}~\bibnamefont
  {Weil}},\ }\href@noop {} {\emph {\bibinfo {title} {Stochastische
  Geometrie}}},\ Teubner Skripten zur mathematischen Stochastik\ (\bibinfo
  {publisher} {B.G. Teubner},\ \bibinfo {year} {2000})\BibitemShut {NoStop}%
\bibitem [{\citenamefont {Schr{\"{o}}der-Turk}\ \emph
  {et~al.}(2010)\citenamefont {Schr{\"{o}}der-Turk}, \citenamefont {Mickel},
  \citenamefont {Schr{\"{o}}ter}, \citenamefont {Delaney}, \citenamefont
  {Saadatfar}, \citenamefont {Senden}, \citenamefont {Mecke},\ and\
  \citenamefont {Aste}}]{EPL10_SchroederTurk}%
  \BibitemOpen
  \bibfield  {author} {\bibinfo {author} {\bibfnamefont {G.~E.}\ \bibnamefont
  {Schr{\"{o}}der-Turk}}, \bibinfo {author} {\bibfnamefont {W.}~\bibnamefont
  {Mickel}}, \bibinfo {author} {\bibfnamefont {M.}~\bibnamefont
  {Schr{\"{o}}ter}}, \bibinfo {author} {\bibfnamefont {G.~W.}\ \bibnamefont
  {Delaney}}, \bibinfo {author} {\bibfnamefont {M.}~\bibnamefont {Saadatfar}},
  \bibinfo {author} {\bibfnamefont {T.~J.}\ \bibnamefont {Senden}}, \bibinfo
  {author} {\bibfnamefont {K.}~\bibnamefont {Mecke}}, \ and\ \bibinfo {author}
  {\bibfnamefont {T.}~\bibnamefont {Aste}},\ }\href@noop {} {\bibfield
  {journal} {\bibinfo  {journal} {Europhysics Lett.}\ }\textbf {\bibinfo
  {volume} {90}},\ \bibinfo {pages} {34001} (\bibinfo {year}
  {2010})}\BibitemShut {NoStop}%
\bibitem [{\citenamefont {Kapfer}\ \emph
  {et~al.}(2012{\natexlab{b}})\citenamefont {Kapfer}, \citenamefont {Mickel},
  \citenamefont {Mecke},\ and\ \citenamefont
  {Schr{\"{o}}der-Turk}}]{PRE12_Kapfer}%
  \BibitemOpen
  \bibfield  {author} {\bibinfo {author} {\bibfnamefont {S.}~\bibnamefont
  {Kapfer}}, \bibinfo {author} {\bibfnamefont {W.}~\bibnamefont {Mickel}},
  \bibinfo {author} {\bibfnamefont {K.}~\bibnamefont {Mecke}}, \ and\ \bibinfo
  {author} {\bibfnamefont {G.}~\bibnamefont {Schr{\"{o}}der-Turk}},\
  }\href@noop {} {\bibfield  {journal} {\bibinfo  {journal} {Physical Review
  E}\ }\textbf {\bibinfo {volume} {85}},\ \bibinfo {pages} {030301(R)}
  (\bibinfo {year} {2012}{\natexlab{b}})}\BibitemShut {NoStop}%
\bibitem [{\citenamefont {Doi}\ and\ \citenamefont
  {Ohta}(1991)}]{DoiOhta:1991}%
  \BibitemOpen
  \bibfield  {author} {\bibinfo {author} {\bibfnamefont {M.}~\bibnamefont
  {Doi}}\ and\ \bibinfo {author} {\bibfnamefont {T.}~\bibnamefont {Ohta}},\
  }\href@noop {} {\bibfield  {journal} {\bibinfo  {journal} {J.~Chem.~Phys.}\
  }\textbf {\bibinfo {volume} {95}},\ \bibinfo {pages} {1242} (\bibinfo {year}
  {1991})}\BibitemShut {NoStop}%
\bibitem [{\citenamefont {Schr{\"o}der-Turk}\ \emph
  {et~al.}(2011{\natexlab{b}})\citenamefont {Schr{\"o}der-Turk}, \citenamefont
  {Trond}, \citenamefont {Campo}, \citenamefont {Kapfer},\ and\ \citenamefont
  {Mickel}}]{Langmuir11_SchroederTurk}%
  \BibitemOpen
  \bibfield  {author} {\bibinfo {author} {\bibfnamefont {G.~E.}\ \bibnamefont
  {Schr{\"o}der-Turk}}, \bibinfo {author} {\bibfnamefont {V.}~\bibnamefont
  {Trond}}, \bibinfo {author} {\bibfnamefont {L.~D.}\ \bibnamefont {Campo}},
  \bibinfo {author} {\bibfnamefont {S.~C.}\ \bibnamefont {Kapfer}}, \ and\
  \bibinfo {author} {\bibfnamefont {W.}~\bibnamefont {Mickel}},\ }\href@noop {}
  {\bibfield  {journal} {\bibinfo  {journal} {Langmuir}\ }\textbf {\bibinfo
  {volume} {27}},\ \bibinfo {pages} {10475} (\bibinfo {year}
  {2011}{\natexlab{b}})}\BibitemShut {NoStop}%
\bibitem [{\citenamefont {Evans}\ \emph
  {et~al.}(2012{\natexlab{a}})\citenamefont {Evans}, \citenamefont
  {Zirkelbach}, \citenamefont {Schr\"oder-Turk}, \citenamefont {Kraynik},\ and\
  \citenamefont {Mecke}}]{Evans:2012}%
  \BibitemOpen
  \bibfield  {author} {\bibinfo {author} {\bibfnamefont {M.~E.}\ \bibnamefont
  {Evans}}, \bibinfo {author} {\bibfnamefont {J.}~\bibnamefont {Zirkelbach}},
  \bibinfo {author} {\bibfnamefont {G.~E.}\ \bibnamefont {Schr\"oder-Turk}},
  \bibinfo {author} {\bibfnamefont {A.~M.}\ \bibnamefont {Kraynik}}, \ and\
  \bibinfo {author} {\bibfnamefont {K.}~\bibnamefont {Mecke}},\ }\href
  {\doibase 10.1103/PhysRevE.85.061401} {\bibfield  {journal} {\bibinfo
  {journal} {Phys. Rev. E}\ }\textbf {\bibinfo {volume} {85}},\ \bibinfo
  {pages} {061401} (\bibinfo {year} {2012}{\natexlab{a}})}\BibitemShut
  {NoStop}%
\bibitem [{\citenamefont {Evans}\ \emph
  {et~al.}(2012{\natexlab{b}})\citenamefont {Evans}, \citenamefont
  {Zirkelbach}, \citenamefont {Schr\"oder-Turk}, \citenamefont {Kraynik},\ and\
  \citenamefont {Mecke}}]{PRE12_Evans}%
  \BibitemOpen
  \bibfield  {author} {\bibinfo {author} {\bibfnamefont {M.~E.}\ \bibnamefont
  {Evans}}, \bibinfo {author} {\bibfnamefont {J.}~\bibnamefont {Zirkelbach}},
  \bibinfo {author} {\bibfnamefont {G.~E.}\ \bibnamefont {Schr\"oder-Turk}},
  \bibinfo {author} {\bibfnamefont {A.~M.}\ \bibnamefont {Kraynik}}, \ and\
  \bibinfo {author} {\bibfnamefont {K.}~\bibnamefont {Mecke}},\ }\href@noop {}
  {\bibfield  {journal} {\bibinfo  {journal} {Physical Review E}\ }\textbf
  {\bibinfo {volume} {85}},\ \bibinfo {pages} {061401} (\bibinfo {year}
  {2012}{\natexlab{b}})}\BibitemShut {NoStop}%
\bibitem [{\citenamefont {Kapfer}\ \emph
  {et~al.}(2012{\natexlab{c}})\citenamefont {Kapfer}, \citenamefont {Mickel},
  \citenamefont {Schr\"{o}der-Turk},\ and\ \citenamefont
  {Mecke}}]{KapferMickelSchroederTurkMecke:2012b}%
  \BibitemOpen
  \bibfield  {author} {\bibinfo {author} {\bibfnamefont {S.}~\bibnamefont
  {Kapfer}}, \bibinfo {author} {\bibfnamefont {W.}~\bibnamefont {Mickel}},
  \bibinfo {author} {\bibfnamefont {G.}~\bibnamefont {Schr\"{o}der-Turk}}, \
  and\ \bibinfo {author} {\bibfnamefont {K.}~\bibnamefont {Mecke}},\
  }\href@noop {} {\enquote {\bibinfo {title} {Spherical minkowski tensors},}\ }
  (\bibinfo {year} {2012}{\natexlab{c}})\BibitemShut {NoStop}%
\bibitem [{\citenamefont {Jerphagnon}, \citenamefont {Chemla},\ and\
  \citenamefont {Bonneville}(1978)}]{AdvancesInPhysics78_Jerphagnon}%
  \BibitemOpen
  \bibfield  {author} {\bibinfo {author} {\bibfnamefont {J.}~\bibnamefont
  {Jerphagnon}}, \bibinfo {author} {\bibfnamefont {D.}~\bibnamefont {Chemla}},
  \ and\ \bibinfo {author} {\bibfnamefont {R.}~\bibnamefont {Bonneville}},\
  }\href@noop {} {\bibfield  {journal} {\bibinfo  {journal} {Adv.~in Phys.}\
  }\textbf {\bibinfo {volume} {27}},\ \bibinfo {pages} {609} (\bibinfo {year}
  {1978})}\BibitemShut {NoStop}%
\bibitem [{\citenamefont {Lautensack}(2007)}]{Lautensack:2007}%
  \BibitemOpen
  \bibfield  {author} {\bibinfo {author} {\bibfnamefont {C.}~\bibnamefont
  {Lautensack}},\ }\href@noop {} {\emph {\bibinfo {title} {Random Laguerre
  Tessellations}}}\ (\bibinfo  {publisher} {Verlag Lautensack, Bingen
  (Germany)},\ \bibinfo {year} {2007})\BibitemShut {NoStop}%
\bibitem [{\citenamefont {Müller}(1953)}]{Mueller:1953}%
  \BibitemOpen
  \bibfield  {author} {\bibinfo {author} {\bibfnamefont {H.}~\bibnamefont
  {Müller}},\ }\href@noop {} {\bibfield  {journal} {\bibinfo  {journal} {Rend
  Circ.~Palermo}\ }\textbf {\bibinfo {volume} {2}} (\bibinfo {year}
  {1953})}\BibitemShut {NoStop}%
\bibitem [{\citenamefont {Lubachevsky}\ and\ \citenamefont
  {Stillinger}(1990)}]{JStatPhys90_Lubachevsky}%
  \BibitemOpen
  \bibfield  {author} {\bibinfo {author} {\bibfnamefont {B.~D.}\ \bibnamefont
  {Lubachevsky}}\ and\ \bibinfo {author} {\bibfnamefont {F.~H.}\ \bibnamefont
  {Stillinger}},\ }\href@noop {} {\bibfield  {journal} {\bibinfo  {journal}
  {J.~Stat.~Phys.}\ }\textbf {\bibinfo {volume} {60}},\ \bibinfo {pages} {561}
  (\bibinfo {year} {1990})}\BibitemShut {NoStop}%
\bibitem [{Note6()}]{Note6}%
  \BibitemOpen
  \bibinfo {note} {In the LS algorithm \cite
  {SkogeDonevStillingerTorquato:2006} spheres are continuously expanded with
  event-driven MD until the pressure exceeds a jamming threshold (jLS). The
  unjammed LS simulations (uLS) used here are stopped at predefined packing
  fractions.}\BibitemShut {Stop}%
\bibitem [{\citenamefont {Skoge}\ \emph {et~al.}(2006)\citenamefont {Skoge},
  \citenamefont {Donev}, \citenamefont {Stillinger},\ and\ \citenamefont
  {Torquato}}]{SkogeDonevStillingerTorquato:2006}%
  \BibitemOpen
  \bibfield  {author} {\bibinfo {author} {\bibfnamefont {M.}~\bibnamefont
  {Skoge}}, \bibinfo {author} {\bibfnamefont {A.}~\bibnamefont {Donev}},
  \bibinfo {author} {\bibfnamefont {F.~H.}\ \bibnamefont {Stillinger}}, \ and\
  \bibinfo {author} {\bibfnamefont {S.}~\bibnamefont {Torquato}},\ }\href@noop
  {} {\bibfield  {journal} {\bibinfo  {journal} {Phys.~Rev.~E.}\ }\textbf
  {\bibinfo {volume} {74}},\ \bibinfo {pages} {041127} (\bibinfo {year}
  {2006})}\BibitemShut {NoStop}%
\bibitem [{Note7()}]{Note7}%
  \BibitemOpen
  \bibinfo {note} {The anisotropy index $\beta _1^{0,2}$ is the ratio of
  eigenvalues of $W_1^{0,2}$ (see Eq.~\protect \textup {\hbox {\mathsurround
  \z@ \protect \normalfont (\ignorespaces \ref {eq:NMT}\unskip \@@italiccorr
  )}}): $\beta _1^{0,2}=\xi _{\protect \mathrm {min}}/\xi _{\protect \mathrm
  {max}}$, where $\xi _{\protect \mathrm {min}}\leq \xi _\protect \mathrm
  {mid}\leq \xi _{\protect \mathrm {max}}$. $\beta _1^{0,2}=1$ indicates
  isotropy, lower values of $\beta _1^{0,2}$ indicate anisotropy \cite
  {EPL10_SchroederTurk}.}\BibitemShut {Stop}%
\bibitem [{\citenamefont {Anikeenko}\ and\ \citenamefont
  {Medvedev}(2007)}]{PRL07_Anikeenko}%
  \BibitemOpen
  \bibfield  {author} {\bibinfo {author} {\bibfnamefont {A.}~\bibnamefont
  {Anikeenko}}\ and\ \bibinfo {author} {\bibfnamefont {N.}~\bibnamefont
  {Medvedev}},\ }\href@noop {} {\bibfield  {journal} {\bibinfo  {journal}
  {Phys.~Rev.~Lett.}\ }\textbf {\bibinfo {volume} {98}},\ \bibinfo {pages}
  {235504} (\bibinfo {year} {2007})}\BibitemShut {NoStop}%
\bibitem [{\citenamefont {Mickel}, \citenamefont {Schröder-Turk},\ and\
  \citenamefont {Mecke}(2012)}]{MickelSchroederTurkMecke:2012}%
  \BibitemOpen
  \bibfield  {author} {\bibinfo {author} {\bibfnamefont {W.}~\bibnamefont
  {Mickel}}, \bibinfo {author} {\bibfnamefont {G.~E.}\ \bibnamefont
  {Schröder-Turk}}, \ and\ \bibinfo {author} {\bibfnamefont {K.}~\bibnamefont
  {Mecke}},\ }\href {\doibase 10.1098/rsfs.2012.0007} {\bibfield  {journal}
  {\bibinfo  {journal} {Interface Focus}\ } (\bibinfo {year} {2012}),\
  10.1098/rsfs.2012.0007}\BibitemShut {NoStop}%
\end{thebibliography}%
\end{document}